\documentclass[fleqn,usenatbib]{mnras}

\usepackage{newtxtext,newtxmath}

\usepackage[T1]{fontenc}

\DeclareRobustCommand{\VAN}[3]{#2}
\let\VANthebibliography\thebibliography
\def\thebibliography{\DeclareRobustCommand{\VAN}[3]{##3}\VANthebibliography}

\usepackage{graphicx}	
\usepackage{amsmath}	

\usepackage{soul} 
\usepackage{amsmath}
\usepackage{xspace}
\usepackage{xifthen}
\usepackage{eso-pic}

\definecolor{forestgreen}{HTML}{228B22}
\definecolor{urlblue}{HTML}{000000}

\newcommand{\Gaia}{{\it Gaia}\xspace}

\mathchardef\mhyphen="2D

\newcommand{\roughly}{\ensuremath{ {\sim}\,} }

\newlength{\dhatheight}

\newcommand{\code}[1]{\texttt{#1}\xspace}

\newcommand{\unit}[1]{\ensuremath{\mathrm{\,#1}}\xspace}

\newcommand{\Gyr}{\unit{Gyr}}
\newcommand{\Myr}{\unit{Myr}}

\newcommand{\pc}{\unit{pc}}
\newcommand{\kpc}{\unit{kpc}}

\newcommand{\Msun}{\unit{M_\odot}}

\newcommand{\secref}[1]{Section~\ref{sec:#1}}

\newcommand{\figref}[1]{Figure~\ref{fig:#1}}

\newcommand{\bandvar}[2][]{%
  \ifthenelse{\isempty{#1}}{\var{#2}}{\var{#2\_#1}}%
}

\newcommand{\LCDM}{\ensuremath{\rm \Lambda CDM}\xspace}

\newcommand{\var}[1]{\ensuremath{\texttt{\MakeUppercase{#1}}}\xspace}

\providecommand\physrep{\ref@jnl{Phys.~Rep.}}%
\providecommand\apjs{\ref@jnl{ApJS}}%
\providecommand{\jcap}{\ref@jnl{JCAP}}%

\usepackage{lineno}

\usepackage{graphicx}
\graphicspath{{./figures/}{./}}

\title[Orbits of streams in Auriga]{Auriga Streams II: orbital properties of tidally disrupting satellites of Milky Way-mass galaxies}

\author[N. Shipp et al.]{Nora Shipp,$^{1}$\thanks{E-mail: nshipp@uw.edu}
Alexander H. Riley,$^{2}$
Christine M. Simpson,$^{3}$
Rebekka Bieri,$^{4}$
Lina Necib,$^{5}$ 
Arpit Arora,$^{1}$\newauthor
Francesca Fragkoudi,$^{2}$
Facundo A.~G\'{o}mez,$^{6}$
Robert J.~J.~Grand,$^{7}$
and Federico Marinacci$^{8,9}$
\\
$^{1}$Department of Astronomy, University of Washington, Seattle, WA 98195, USA\\
$^{2}$Institute for Computational Cosmology, Department of Physics, Durham University, South Road, Durham DH1 3LE, UK\\
$^{3}$Argonne Leadership Computing Facility, Argonne National Laboratory, Lemont, IL 60439, USA\\
$^{4}$Department of Astrophysics, University of Zurich, 8057 Zurich, Switzerland\\
$^{5}$Department of Physics and Kavli Institute for Astrophysics and Space Research, Massachusetts Institute of Technology, 77 Massachusetts Ave, Cambridge,\\ MA 02139, USA\\
$^{6}$Departamento de Astronom\'{i}a, Universidad de La Serena, Av.~Juan Cisternas 1200 Norte, La Serena, Chile\\
$^{7}$Astrophysics Research Institute, Liverpool John Moores University, 146 Brownlow Hill, Liverpool, L3 5RF, UK\\
$^{8}$Department of Physics \& Astronomy `Augusto Righi', University of Bologna, via Gobetti 93/2, I-40129 Bologna, Italy\\
$^{9}$INAF, Astrophysics and Space Science Observatory Bologna, Via P. Gobetti 93/3, I-40129 Bologna, Italy
}

\date{Accepted XXX. Received YYY; in original form ZZZ}

\pubyear{2024}

\begin{document}
\label{firstpage}
\pagerange{\pageref{firstpage}--\pageref{lastpage}}
\maketitle

\begin{abstract}
Galaxies like the Milky Way are surrounded by complex populations of satellites at all stages of tidal disruption. In this paper, we present a dynamical study of the disrupting satellite galaxies in the Auriga simulations that are orbiting 28 distinct Milky Way-mass hosts across three resolutions. We find that the satellite galaxy populations are highly disrupted. The majority of satellites that remain fully intact at present day were accreted recently without experiencing more than one pericentre ($n_{\rm peri} \lesssim 1$) and have large apocentres ($r_{\rm apo} \gtrsim 200 \kpc$) and pericentres ($r_{\rm peri} \gtrsim 50 \kpc$). The remaining satellites have experienced significant tidal disruption and, given full knowledge of the system, would be classified as stellar streams. We find stellar streams in Auriga across the range of pericentres and apocentres of the known Milky Way dwarf galaxy streams and, interestingly, overlapping significantly with the Milky Way intact satellite population. We find no significant change in satellite orbital distributions across resolution. However, we do see substantial halo-to-halo variance of $(r_\text{peri}, r_\text{apo})$ distributions across host galaxies, as well as a dependence of satellite orbits on host halo mass -- systems disrupt at larger pericentres and apocentres in more massive hosts. Our results suggest that either cosmological simulations (including, but not limited to, Auriga) are disrupting satellites far too readily, or that the Milky Way's satellites are more disrupted than current imaging surveys have revealed. Future observing facilities and careful mock observations of these systems will be key to revealing the nature of this apparent discrepancy.
\end{abstract}




\section{Introduction}

\begin{figure*}
    \includegraphics[width=1.0\textwidth]{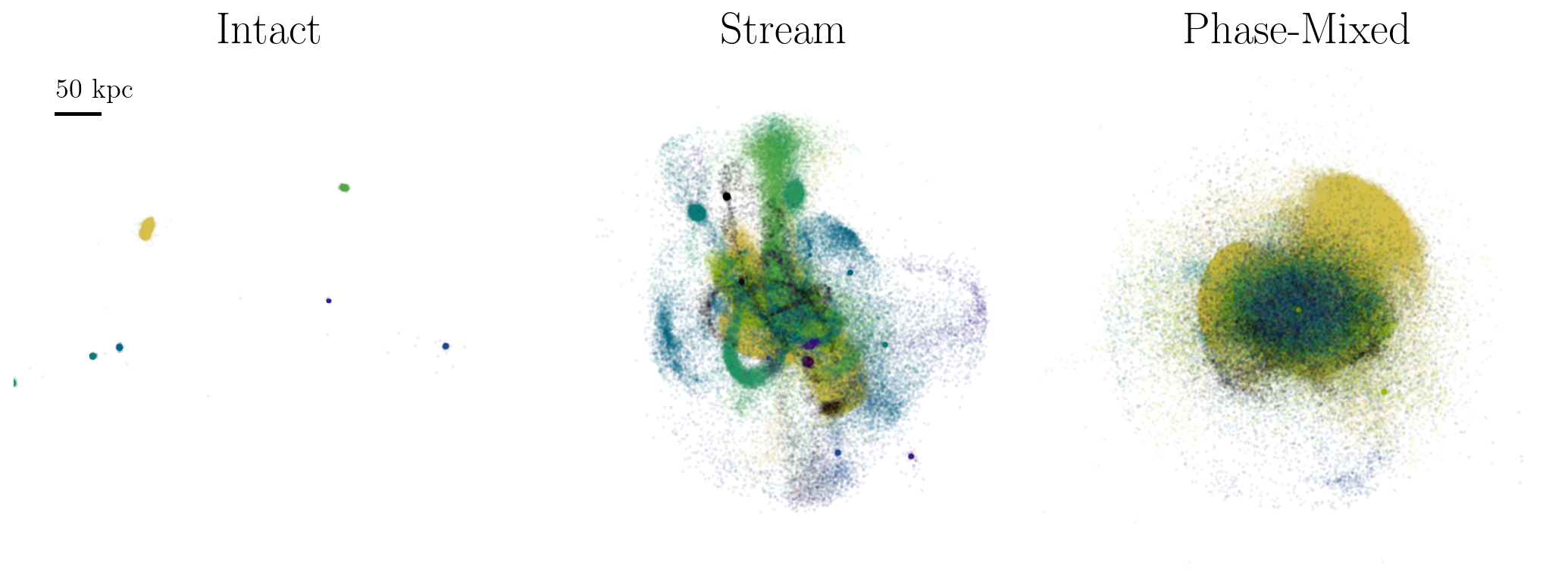}
    \caption{Complete population of accreted satellites, including intact satellites (left), streams (middle), and phase-mixed systems (right) around one of the Auriga haloes (Au-23, level 3), shown in the simulation's Cartesian coordinate system centred on the host galaxy. Only a small fraction of accreted systems remain intact and many of the surviving progenitors have extended tidal tails and are classified as stellar streams. The colors are assigned to each accreted satellite arbitrarily in order to differentiate the member stars between systems. An animation of the positions of these systems over time can be found at \href{https://www.norashipp.com/s/auriga_streams_2_figure_1.gif}{this url.}}
    \label{fig:intro}
\end{figure*}

In the standard $\Lambda$ Cold Dark Matter (\LCDM) paradigm, galaxies like the Milky Way (MW) experience a series of merger and accretion events \citep{Searle:1978}. Many of these accreted systems are less massive dwarf galaxies, which first orbit the host galaxy before being disrupted by the tidal forces of the host's gravitational potential and ultimately phase-mix into a smooth stellar halo \citep{Bullock:2005}. We therefore expect galaxies like our own to be surrounded by populations of \textit{intact satellite galaxies,} systems currently undergoing tidal disruption such as extended \textit{stellar streams,} and a smooth stellar halo consisting of the \textit{phase-mixed} remnants of past accretion events.

Observations of the MW's halo have indeed revealed such structures. In particular, large, wide-area photometric surveys including the Sloan Digital Sky Survey \citep[SDSS;][]{York:2000}, the Dark Energy Survey \citep[DES;][]{DES:2016}, and other DECam surveys \citep[][]{Drlica-Wagner:2021}, have enabled the discovery of $>50$ surviving satellite galaxies around the MW \citep[e.g.][]{Drlica-Wagner:2020}. These surveys, as well as the all-sky astrometric survey by the \Gaia satellite, have also revealed large numbers of tidally disrupting stellar streams \citep[e.g.][]{Belokurov:2006, Grillmair:2006, Koposov:2014, Bernard:2016, Malhan:2018a, Shipp:2018}, with the total number of stream candidates now exceeding 100 \citep{Mateu:2023, BonacaPriceWhelan:2024}. The majority of these stellar streams are confirmed or believed to have globular cluster progenitors, with only eight distinct streams confirmed to originate from dwarf galaxies \citep{Li:2022}. The unprecedented proper motion measurements of MW stars by \Gaia, as well as radial velocity measurements from spectroscopic surveys, has also revealed an abundance of more phase-mixed structures in the MW's stellar halo, most notably the \Gaia-Sausage-Enceledus (GSE) merger \citep{Helmi:2018, Belokurov:2018, Naidu:2020}. Furthermore, these observations of the kinematics of stars in the MW's halo have enabled unprecedented dynamical characterisation and modeling of these accreted systems \citep[e.g.][]{Kallivayalil:2018, Patel:2020, Erkal:2019, Shipp:2021, Pace:2022, Koposov:2023}.

Additionally, these rich datasets have exposed the complexity of these systems, and the difficultly in separating out truly intact from disrupting satellites. For example, many intact satellites have been found to have stars beyond their tidal radii \citep[e.g.][]{Carlin:2018, Li:2021, Chiti:2021, Filion:2021, Ji:2021, Qi:2022, Sestito:2023, Ou:2024hercules, Jensen:2024}, although in many cases the cause of these extra-tidal stars remains uncertain, with explanations including tidal disruption due to the main host \citep{Fattahi:2018, Ou:2024hercules}, accretion from smaller systems \citep{Tarumi:2021, Deason:2022}, internal feedback mechanisms \citep{Revaz:2018}, and alternative models of dark matter \citep{Pozo:2024}. Comparisons between the observed MW accreted satellite population and cosmological simulations of MW-like galaxies may help to identify the origin of these extended stellar distributions, and to broaden our understanding of the full population of satellites around our own Galaxy.

Studies of intact satellite galaxy populations across simulations have long produced important results on the physics of small-scale galaxy formation \citep[e.g.][]{Brooks:2014, Wetzel:2016, Sawala:2016, Garrison-Kimmel:2018, Grand:2021, Munshi:2021}, constraints on the properties of dark matter \citep[e.g.][]{Bullock:2017, Nadler:2021, Sales:2022}, and insight into the formation history of our own Galaxy \citep[e.g.][]{Rocha:2012, Gomez:2013, Fattahi:2019, Fattahi:2020, Bose:2020, Samuel:2020, Vera-Casanova:2022}. Only recently have hydrodynamic cosmological simulations reached resolutions that enable similar population-level studies of disrupting satellites around MW-mass hosts. \citet{Panithanpaisal:2021} and \citet{Shipp:2023} studied populations of stellar streams in the Feedback in Realistic Environments (FIRE-2) cosmological simulations. They classified accreted systems around 13 MW-mass host galaxies as intact dwarf galaxies, coherent stellar streams, or phase-mixed systems and produced synthetic observations of the simulated data in order to make comparisons to the observed MW stellar streams. They found that the number and stellar mass function of detectable streams around the FIRE-2 galaxies are consistent with those observed around the MW. However, they identified a discrepancy in the orbits of these systems, finding that even the \textit{detectable} stellar streams in FIRE-2 disrupted at larger pericentres and apocentres than those in the MW, and that nearly all of the FIRE-2 satellites on orbits consistent with the MW streams have been fully phase-mixed. In addition, they found that many systems that have extended tidal tails would only be detected as intact satellites given the surface brightness limits of current surveys like DES. These conclusions raise the question of whether satellite galaxies are disrupting at artificially high rates in simulations, or whether the MW satellite galaxy population may in fact be more disrupted than previously observed. Distinguishing between these possibilities requires studies of disrupting satellites across simulations with a range of simulation codes, sub-grid physics models, host galaxy properties, and simulation resolutions.

In this work, we present a dynamical study of disrupting satellites in the Auriga cosmological simulations. We derive the orbital properties of the full population of accreted satellite galaxies, at all stages of tidal disruption, across 28 simulated MW-mass hosts at three different resolutions. In \secref{sims} we introduce the Auriga simulations used in this work, as well as their populations of intact satellite galaxies, stellar streams, and phase-mixed systems. In \secref{results} we present the orbital properties of these systems, and discuss how orbits vary with satellite and host properties as well as simulation resolution. In \secref{discussion} we summarise our conclusions and discuss the implications of the results for populations of disrupting satellites across simulations and in the MW.
This paper accompanies Riley et al.~(2024), hereafter Paper I, which presents the catalogue of accretion events in Auriga.

\section{Disrupting Satellites in Auriga}
\label{sec:sims}

\subsection{Auriga Simulations}
\label{sec:auriga}

The Auriga project \citep{Grand:2017, Grand:2024} consists of cosmological, magneto-hydrodynamic zoom-in simulations of 30 MW-mass galaxies (Au-1 to Au-30). These galaxies have halo masses ($M_\text{200c}$\footnote{Defined to be the mass enclosed in a sphere in which the mean matter density is 200 times the critical density $\rho_\text{crit} = 3H^2(z) / 8\pi G$. Virial quantities are defined at this radius and identified with a `200c' subscript.}) within the range $1 -2 \times 10^{12} \Msun$ and have a range of accretion histories and other halo and galaxy properties. These simulations were carried out using the moving-mesh code \textsc{Arepo} \citep{Springel:2010, Pakmor:2016} and include a comprehensive galaxy formation model that incorporates primordial and metal-line cooling \citep{Vogelsberger:2013}, a spatially-uniform redshift-dependent UV background for reionization \citep{Faucher-Giguere:2009}, star formation and feedback mechanisms \citep{Springel:2003, Vogelsberger:2013}, magnetic fields \citep{Pakmor:2017}, and the modeling of black holes, including seeding, accretion, and feedback \citep{Springel:2005feedback, Marinacci:2014, Grand:2017}.

The Auriga haloes were selected from the EAGLE dark-matter-only simulation box with a co-moving side length of 100 Mpc \citep{Schaye:2015}, chosen not only for their specific mass range but also for their relative isolation. The simulations use the cosmological parameters from \citet{PlanckCollaboration:2014}, with $\Omega_\text{M} = 0.307$, $\Omega_\Lambda = 0.693$, $h = 0.6777$, $\sigma_8 = 0.8288$, and $n_s = 0.9611$. Initial conditions were generated using \textsc{Panphasia} \citep{Jenkins:2013}.

In this work we use three different resolution levels of these simulations. We consider 28 haloes that have  been simulated at level 4\footnote{We exclude Au-1 and Au-11 because they are undergoing massive accretion events at $z=0$.}, with dark matter particle mass, baryonic particle mass, and minimum particle softening length equal to $m_{\rm DM} = 3 \times 10^5 \Msun,\ m_{\rm b} = 5 \times 10^4 \Msun,\ h_{\rm b} = 375 \pc$. Six haloes (Au-6, Au-16, Au-21, Au-23, Au-24, Au-27) have been resimulated at level 3 ($m_{\rm DM} = 4 \times 10^4 \Msun, m_{\rm b} = 6 \times 10^3 \Msun, h_{\rm b} = 188 \pc$), and one halo (Au-6) has been resimulated at level 2 \citep[$m_{\rm DM} = 4,600 \Msun, m_{\rm b} = 850 \Msun, h_{\rm b} = 94 \pc$;][]{Grand:2021}. We consider level 3 to be our fiducial resolution level. This mass resolution is roughly equivalent to that of the FIRE-2 \textit{Latte} simulations \citep{Wetzel:2016} used by \citet{Shipp:2023}, enabling studies of disrupting satellites down to $M_* \roughly 5 \times 10^5 \Msun$, and strikes a balance between halo sample size and simulation resolution. We present a study of satellite orbits across all three resolution levels in \secref{resolution}, and consider the larger sample of level 4 simulations to study the effect of host galaxy properties on satellite orbits in \secref{hosts}.

Halo catalogs in the Auriga simulations are generated using the on-the-fly SUBFIND halo finder \citep{Springel:2001}. This process begins with a Friends-of-Friends (FOF) algorithm that groups dark matter particles into potential haloes based on a standard linking length criterion \citep{Davis:1985}. Once these haloes are identified, the SUBFIND algorithm further refines the results by separating each halo into gravitationally self-bound subhaloes, ensuring that the catalog accurately reflects the complex structure of these systems. The FOF algorithm initially groups only dark matter particles, while other particle types, such as gas, stars, and black holes, are assigned to the nearest dark matter group. SUBFIND is then applied across all particle types within each FOF group simultaneously. The merger trees are then constructed in post-processing with the LHaloTree algorithm \citep{Springel:2005}, enabling precise tracking of individual haloes, subhaloes, and their interactions over cosmic time. This method effectively traces the assembly and growth of galaxies within the Auriga framework.

The Auriga simulations accurately capture observed characteristics of spiral galaxies, including stellar masses, sizes, rotation curves \citep{Grand:2017}, magnetic fields \citep{Pakmor:2017}, and the distribution of neutral hydrogen (H\textsc{i}; \citealp{Marinacci:2017}). Moreover, the Auriga simulations have played a crucial role in investigating the structure and dynamics of satellite galaxy systems \citep{Simpson:2018} and stellar haloes \citep{Monachesi:2016, Monachesi:2019, Vera-Casanova:2022}, as well as offering insights into the MW’s assembly history \citep{Deason:2017, Fattahi:2019}. The Auriga simulations have also contributed to understanding the orbits of the MW's satellites \citep{Riley:2019} and have been used to estimate the mass of the MW, producing results that align well with a range of observational constraints \citep{Callingham:2019, Deason:2019}. We build our analysis on the foundation of these works, with a particular focus on the results exploring satellite dynamics and their interactions with the host galaxy.

\begin{figure*}
    \includegraphics[width=1.0\textwidth]{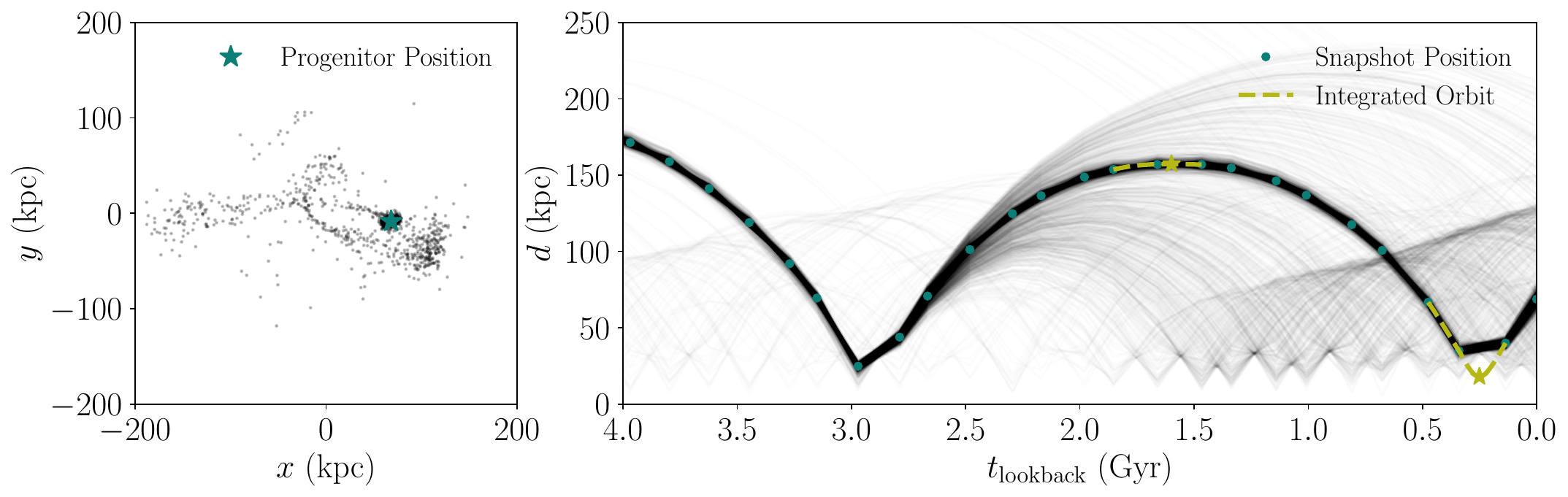}
    \caption{Example stream orbit demonstrating the process of calculating pericentres and apocentres of simulated satellites. The left panel shows the spatial distribution of the member stars of a simulated stellar stream, and the high-density progenitor region is marked by a blue star. On the right, the distance as a function of look-back time is shown for each member star (black lines). The blue points indicate the median position of the selected progenitor stars from the simulation snapshots. The yellow curves show the integrated orbits around the pericentre and apocentre and the yellow stars show the calculated pericentre and apocentre values.}
    \label{fig:orbit_fit}
\end{figure*}

\subsection{Catalog of Disrupting Satellites}
\label{sec:streams}

The Auriga catalog of disrupting satellites is presented in detail in Paper I. In short, we first identify all of the accretion events across the history of each of the 28 simulated host galaxies. We then identify all of the simulation star particles associated with each accretion event (regardless of whether they are currently bound to the accreted system) and determine their positions and velocities at $z=0$. Due to resolution limits, we consider all accretions with $>100$ associated star particles. This corresponds to $M_* \roughly 5 \times 10^6,\ 6 \times 10^5,$ and $8.5 \times 10^4 \Msun$ at levels 4, 3, and 2, respectively. 

We then classify systems as \textit{intact satellite}, \textit{stellar stream}, or \textit{phase-mixed}. We determine whether a satellite remains intact based on the fraction of associated stars that are bound to the satellite progenitor at $z=0$. Intact satellites are defined as systems with $f_{\rm bound} > 0.97$. We differentiate between coherent streams and phase-mixed structures based on the median local velocity dispersion ($\sigma_{\rm 50}$) of the star particles associated with each system. Systems are classified as phase-mixed if $\sigma_{\rm 50} > 3.51 \log \left( \frac{M_\star}{M_\odot} \right) + 1.08$, where the coefficients are fit via a support vector machine model applied to a visually classified sample. This technique was first used to classify tidal debris as coherent stream or phase-mixed in the FIRE-2 simulations by \citet{Panithanpaisal:2021}, and the method as applied to Auriga is described in greater detail in Paper I\footnote{In this classification scheme, shells -- disrupting systems that may be extended both parallel and perpendicular to their progenitor orbit -- would primarily be classified as phase-mixed.}.
An important point to note is that we have made no correction for the detectability of the Auriga systems; as in \citet{Shipp:2023}, many systems with tidal debris are likely too faint to detect, or would only be detected as intact progenitors given current imaging.

Typically we find that the majority of accretion events that we catalogue have been phase-mixed, followed by a significant fraction of systems ($\roughly 30\%$) that are currently undergoing tidal disruption and are classified as stellar streams, and a relatively small fraction of accretions ($\lesssim 10\%$) that survive to $z=0$ as intact satellites. This can be seen in \figref{intro}, which illustrates the complete accreted satellite population around one of the simulated Auriga haloes (Au-23), divided by morphological classification. \textit{Notably, the majority of systems with surviving progenitors also have extended tidal tails and are classified as stellar streams.}

\section{Orbital Properties of Disrupting Satellites}
\label{sec:results}

\subsection{Orbit Fitting}
\label{sec:fitting}

To determine the orbital properties of each disrupting satellite, we trace the positions of each member star particle back through the output simulation snapshots. The frequency of snapshots varies across resolution level. Near $z=0$ the snapshot spacing is $\roughly 130 \Myr$ for level 4, $\roughly 330 \Myr$ for level 3, and for level 2 the star particles are output at a higher frequency of $\roughly 10 \Myr$. The level 2 snapshots are saved at a high enough frequency to reliably recover the pericentres and apocentres of the system. However, for both level 4 and level 3 we integrate the orbits between snapshots in order to ensure we are calculating the pericentre and apocentre to sufficient accuracy. 

To do this, we use \code{SUBORBITIN} \citep{Richings:2020} and \code{AGAMA} \citep{Vasiliev:2019} to model the potential of the host galaxy near each pericentre snapshot and to integrate the satellites orbits. Following the well-established method of simulation orbit reconstruction using basis function expansions \citep{Sanders:2020, Arora:2022, Arora:2024}, we describe the host galaxy potential via a multipole expansion with $l_{\rm max} = 4$, using a combination of spherical harmonics and azimuthal harmonics for the halo and
disc respectively. We then compute a weighted average of the potential of the two snapshots nearest to the calculated pericentre or apocentre and integrate the orbit in this combined potential. One simulated galaxy (Au-18) was resimulated with an ultra-high frequency snapshot cadence of  $\roughly 5 \Myr$. We test our orbit fitting method on this simulation and confirm that we can indeed accurately recover even small pericentres at the snapshot frequencies of the level 4, 3, and 2 simulations.

We define the pericentre (apocentre) as the most recent closest (farthest) approach to the centre of the host galaxy of the progenitor orbit. This is most analogous to what is measured for observed systems, and consistent with the values presented for the FIRE-2 simulations in \citet{Shipp:2023}. For intact satellites and stellar streams, we identify the $z=0$ progenitor location by fitting a 3D Gaussian kernel density estimate (KDE) to the spatial coordinates of the member stars. We select stars residing within the highest density region, take the median of their position and velocity coordinates (which we consider to represent the progenitor location) and integrate the progenitor orbit using the above method. By first calculating the progenitor position at each snapshot and then considering the bulk progenitor orbit rather than the orbit of each individual member star, in addition to interpolating between snapshots using the weighted averaging framework of \citet{Richings:2020}, we eliminate the need to consider the self-gravity of the progenitor during the integration. For phase-mixed systems as well as a small number of streams, which no longer have a high-density region corresponding to the progenitor location, we integrate the orbits of each individual member star, then take the median pericentre and apocentre values.

A small subset of systems have not had a pericentre and/or apocentre after accretion, where accretion is defined as the first time of crossing $R_{\rm 200c}$ of the host galaxy. In that case we consider their current distance to be an upper limit on the pericentre and lower limit on the apocentre. These values are plotted as upper/lower limits in any figures where they are included.

\figref{orbit_fit} illustrates the orbit fitting procedure for one example stream. The left-hand panel shows the spatial coordinates of the stream member stars at $z=0$, with the selected progenitor location highlighted with a blue star. The right-hand panel shows the distance of each member star from the centre of the host galaxy over time. The stars in the high-density progenitor region can be identified as the thick black line from which other orbits diverge over time. The snapshot progenitor positions are plotted as blue points and the integrated orbits around pericentre and apocentre are shown in yellow. The final pericentre and apocentre values are indicated with yellow stars.
This example highlights the need to model orbits where snapshot cadences are sparse - using only the snapshot positions of the star particles would give a most recent pericentre of $\sim$40~kpc, while interpolating between snapshots yields $\sim$20~kpc.

\subsection{Satellite Orbits}
\label{sec:orbits}

\begin{figure*}
    \includegraphics[width=1.0\textwidth]{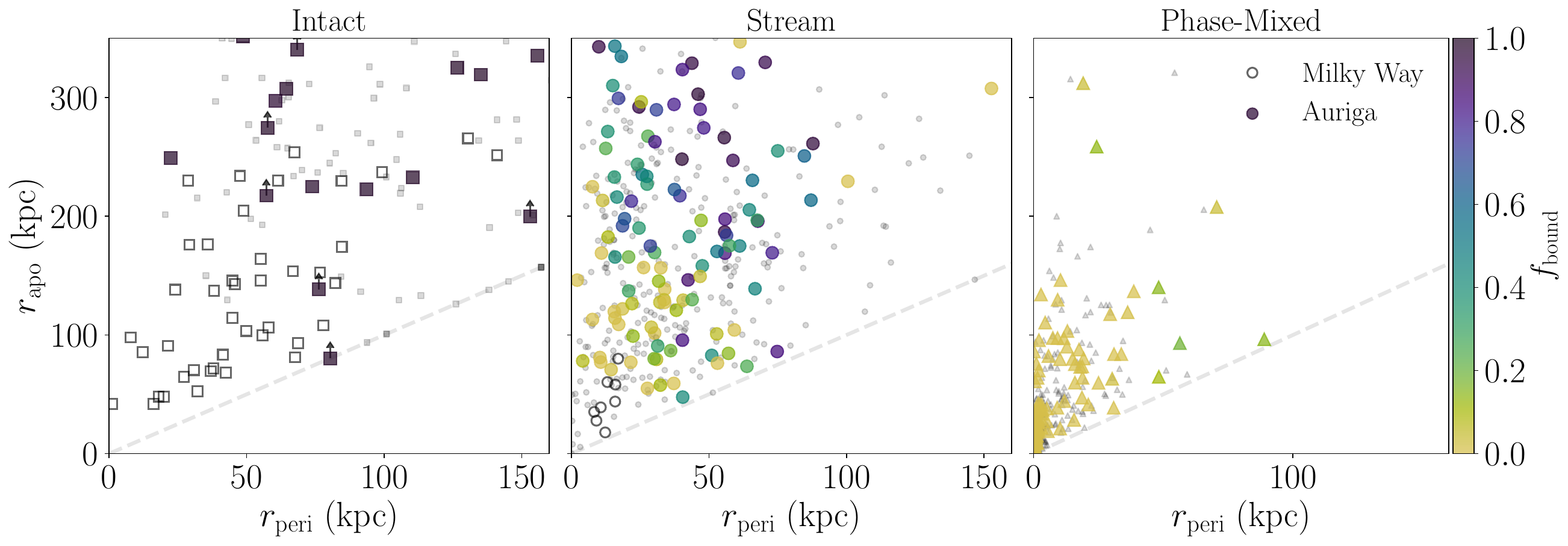}
    \caption{Pericentres and apocentres of intact satellites (left panel), stellar streams (middle panel), and phase-mixed systems (right panel) in the Auriga simulations. The fiducial level 3 simulations are colored by the fraction of associated stars that remain bound to the progenitor at $z=0$ (by definition, all intact systems have $f_\text{bound}\sim 1$). Other resolutions (level 4 and level 2) are plotted in the background as light gray points to illustrate the variation across the full sample of 28 simulated host galaxies. The dashed gray line indicates circular orbits (equal pericentre and apocentre). The unfilled markers represent the MW satellites (squares) and streams (circles) for comparison, see Section \ref{sec:orbits} for data sources. Note that there is no completeness or detectability correction applied here, so some of the Auriga systems may not be detectable given current observations.
    }
    \label{fig:peri_apo_l3}
\end{figure*}

\begin{figure}
    \includegraphics[width=1.0\columnwidth]{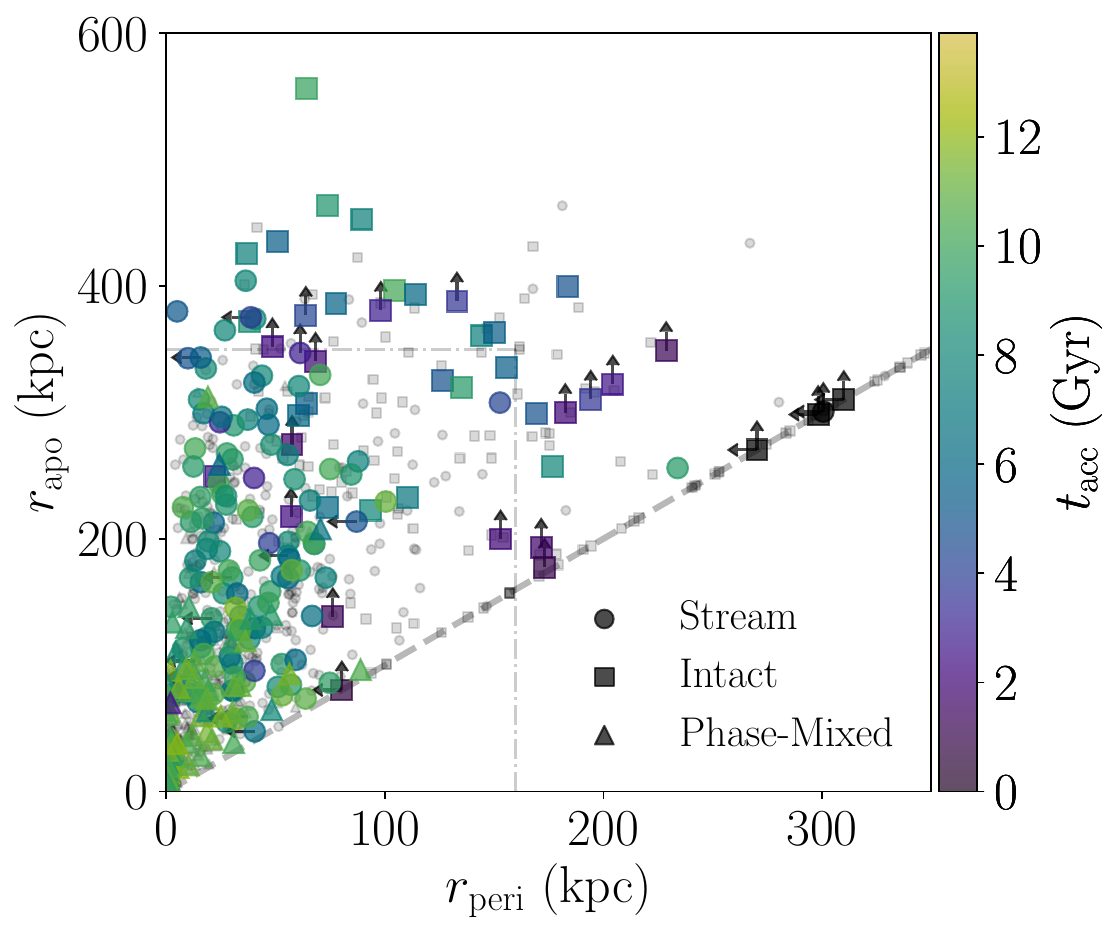}
    \caption{Full range of pericentres and apocentres of intact satellites (squares), stellar streams (circles), and phase-mixed systems (triangles) in Auriga. The fiducial level 3 simulation points are colored by the look-back time to accretion of each system. The level 4 and level 2 resolution points are included as light gray points in the background. The thick gray-dashed line indicates circular orbits, and the light dashed gray box indicates the axis limits shown in \figref{peri_apo_l3}. At large pericentres and apocentres there are several intact satellites on first infall, that have not had a pericentre or apocentre since accretion and are therefore shown with their current distance as an upper/lower limit.}
    \label{fig:peri_apo_l3_full}
\end{figure}

\begin{figure}
    \includegraphics[width=0.95\columnwidth]{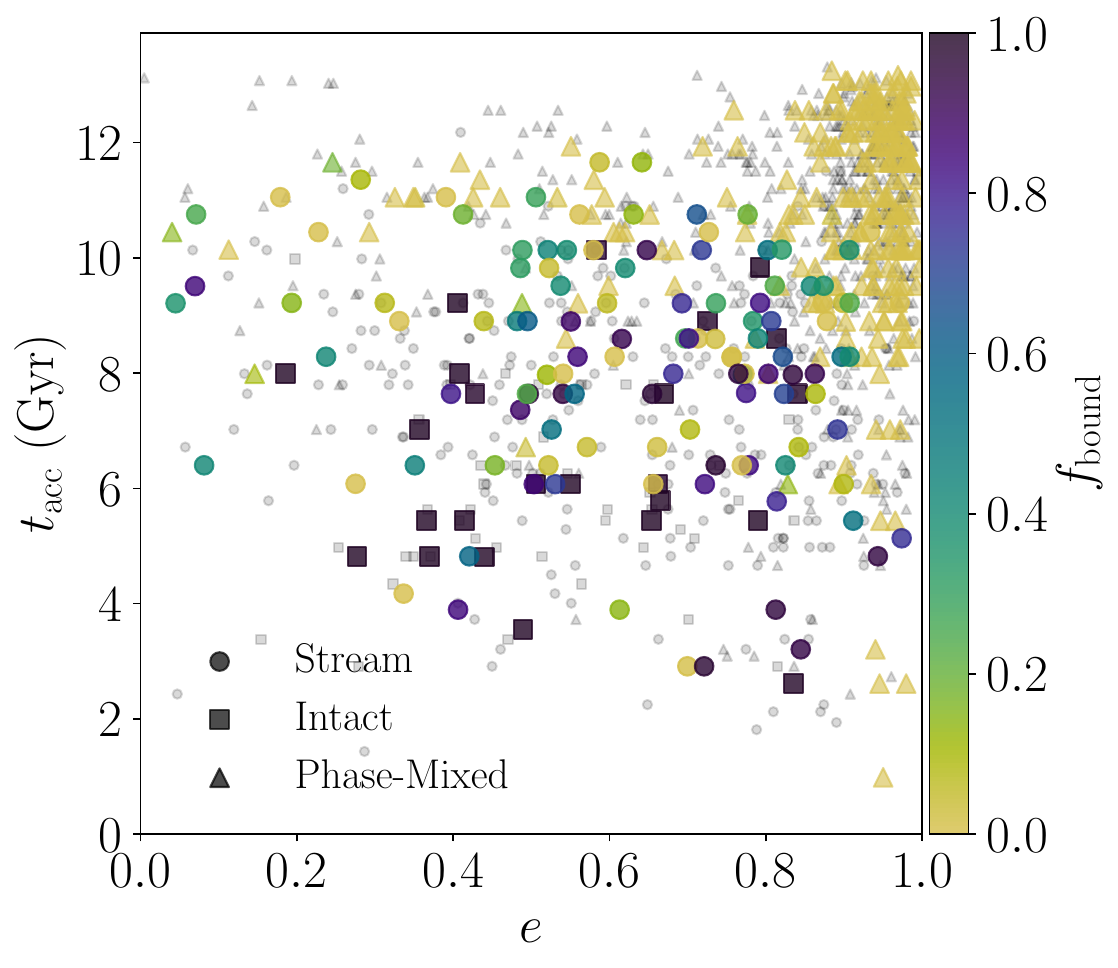}
    \caption{Look-back time to accretion versus orbital eccentricity of disrupting satellites in Auriga. As in the previous figures, the fiducial level 3 simulation points are colored by the stellar bound fraction of each system, while the other resolution points are included in the background. Phase-mixed systems (triangles) nearly all follow very eccentric orbits and/or were accreted $> 10 \Gyr$ ago. Streams and intact satellites follow a broader distribution, however intact satellites generally have been accreted more recently, and if not tend to be on more radial orbits with very large apocentres.}
    \label{fig:tacc_ecc}
\end{figure}

\figref{peri_apo_l3} presents the pericentres and apocentres of intact satellites, stellar streams, and phase-mixed systems across the Auriga simulations. Systems from the fiducial level 3 simulations are shown as colored points, where the color corresponds to the fraction of stars that remain bound at $z=0$. Level 2 and 4 systems are included as gray background points to demonstrate the broad range across the 28 unique host galaxies. The left panel shows the intact satellites (squares). The unfilled boxes indicate the MW satellite orbits from \citet{Pace:2022}. These orbits are calculated in a static three-component MW potential, plus an infalling LMC and the resulting reflex motion of the centroid of the MW. We assign satellites with confirmed tidal tails (Sagittarius and Tucana III) as streams, but include satellites such as Crater II and Antlia II which have proposed signs of tidal disruption \citep{Torrealba:2019, Ji:2021, Pace:2022, Vivas:2022} as intact. It is highly likely that some of the simulated systems would not yet be detectable in our Galaxy given available data. We leave mock observations of the Auriga streams for future work. Note that several intact satellites are recently accreted and have not had a pericentre and/or apocentre after accretion and are therefore plotted with their current distance as an upper/lower limit, as discussed in \secref{fitting}. 

The middle panel shows the systems classified as stellar streams (circles). The unfilled points represent the MW dwarf galaxy streams. The pericentre and apocentre values are taken from \citet{Li:2022}. As for the intact satellites, the stream orbits are calculated in a MW potential that includes the effect of an infalling constant-mass LMC and the reflex motion of the MW, but does not include any other time dependence of the potential. The right panel shows the phase-mixed systems (triangles). No MW systems are plotted in this case, because of the difficulty in measuring orbital parameters for observed phase-mixed structures. We note that the pericentres and apocentres for the phase-mixed systems in Auriga are generally less informative than for the streams and satellites due to the large spread in orbits among member stars and the impossibility of selecting only the progenitor orbit \citep{Khoperskov:2023, Mori:2024}. The axis limits of the figure are limited to highlight the range overlapping with the MW systems. 

The full extent to the pericentre-apocentre distributions is shown in \figref{peri_apo_l3_full}, where the level 3 points are colored by lookback time to accretion and the level 4 and 2 systems are again plotted as background gray points. Importantly, we find that the Auriga simulations do include several cases of streams and even phase-mixed systems forming around satellites, which can lead to deceptively large pericentres and apocentres for disrupted systems. This affects the overall pericentre and apocentre distributions, but largely only contributes significant outliers that can be seen in \figref{peri_apo_l3_full} and does not bias the distribution shown in \figref{peri_apo_l3}. The effect of satellites-of-satellites will be discussed in greater detail in \secref{satsofsats}.

\figref{tacc_ecc} shows the accretion time and eccentricity of systems classified as intact (squares), streams (circles), and phase-mixed (triangles) colored by their stellar bound fraction. This figure includes only systems that have had a pericentre and apocentre since accretion and therefore have a well-defined orbital eccentricity. All of these figures are discussed in greater detail in the following subsections.

\subsubsection{Intact Satellites}
\label{sec:intact_orbs}

Intact satellites in Auriga tend to be orbiting at large apocentres, fairly large pericentres, and to have completed a small number of orbits since accretion. In particular, all intact satellites (that have a calculated apocentre) have $r_{\rm apo} > 200 \kpc$, in contrast with the large population of MW satellites at much smaller apocentres. In addition, with some exceptions, the intact Auriga satellites have $r_{\rm peri} > 50 \kpc$, again in apparent disagreement with observed MW satellites. By definition, the intact satellites have large values of $f_{\rm bound} > 0.97$. In \figref{peri_apo_l3_full} and \figref{tacc_ecc}, it can be seen that the majority of intact satellites, particularly if they have smaller pericentres or apocentres, were accreted recently ($t_{\rm acc} < 6 \Gyr$). Earlier-accreted intact systems tend to be on more radial orbits ($e > 0.5$) and therefore have large apocentres and have spent less time in the inner galaxy (the few with $e < 0.5$ have large pericentres). In fact, we find that most intact satellites have not experienced more than one pericentre (across the level 3 simulations, $\roughly 25\%$ have had zero pericentres, $\roughly 60\%$ have had one, and $\roughly 15\%$ have had two).

There is an apparent discrepancy between the orbits of the Auriga intact satellites and those of the observed MW satellites, which could be due to a range of observational, theoretical, or numerical effects.
On the theoretical side, there are both uncertainties when computing the MW satellite orbits \citep{Santistevan:2024, D'Souza:2022} and potential over-disruption in the simulations (due to numerics or galaxy formation or dark matter models). If the simulations are correct, then it could be that the `intact' MW satellite population is in fact more disrupted than has so far been observed and would be classified as streams under our framework (if they have lost more than $3\%$ of their stellar mass).

\citet{Shipp:2023} showed that the uncertainty due to the fact that the MW satellite orbits are integrated in a largely time-independent potential is negligible relative to the scale of the differences, suggesting that uncertainty on measured orbital parameters cannot account for the full discrepancy. 
Properties of the simulations and the possibility of over-disruption will be discussed in greater detail in \secref{discussion}. 
A key remaining uncertainty lies in the detectability of the Auriga satellites. 
\citet{Shipp:2023} found that in the FIRE-2 \textit{Latte} simulations, many disrupting satellites would be misclassified as intact in DES-depth data, given the low surface brightness of their tidal tails relative to the high-density progenitor. This is likely true for many of the Auriga satellites as well and will be addressed in future work.
The apparent discrepancy between the orbits of intact satellites in Auriga and the MW will be discussed in greater detail in \secref{discussion}.

\subsubsection{Stellar Streams}
\label{sec:stream_orbs}

The Auriga stellar streams span the broadest range of pericentre-apocentre space, extending from very small values ($r_{\rm peri}, r_{\rm apo} < 10 \,\kpc$) to, in rare cases, $r_{\rm peri} > 100 \,\kpc$, $r_{\rm apo} > 300 \,\kpc$. There is significant variation in $f_{\rm bound}$ between values of 0 and 0.97 and a strong correlation with orbital radius, with streams with $r_{\rm peri} \lesssim 50 \,\kpc$ and $r_{\rm apo} \lesssim 150 \,\kpc$ having lost nearly all of their stars. In Auriga, only $\roughly 30\%$ of stellar streams have $f_{\rm bound} < 5\%$, whereas the majority of the known MW dwarf galaxy streams ($>80\%$, with only the exception of Sagittarius and Tucana III) do not have known associated progenitors. Given the small pericentres and apocentres of the MW streams, this picture is not inconsistent with the low $f_{\rm bound}$ of Auriga streams on similar orbits.

However, there is a notable difference between the apparent orbital distributions of the simulated and observed stellar streams, with streams forming in the Auriga simulations at much larger pericentres and apocentres than we have yet to observe in the MW. 
Furthermore, the majority of Auriga systems on orbits consistent with the surviving MW satellite population are classified as stellar streams and have lost a significant fraction of their stellar mass ($f_{\rm bound} \lesssim 0.8$). 
As mentioned above, it is important to keep in mind that we have made no completeness correction to the MW distribution, nor have we evaluated the detectability of the simulated stellar streams. It is possible that, after conducting mock observations, many of the Auriga streams would be misclassified as intact satellites that have experienced no disruption, given current observational capabilities.

In the FIRE-2 simulations, \citet{Shipp:2023} observed that accreted satellites on orbits consistent with the MW dwarf galaxy streams are almost entirely phase-mixed. This is also true for the limited sample of six resolution level 3 simulations, but is not true across all resolution levels. In \figref{peri_apo_l3}, it can be seen that there are several systems among the full sample of 28 haloes (gray points) with smaller pericentre and apocentre values that are consistent with the MW observed streams. In particular, 4 streams from Au-7 and 3 streams from Au-(4, 5, 15, 21, 22) have $r_\text{peri} < 25$~kpc and $r_\text{apo} < 100$~kpc (note that the minimum stellar mass we consider in level~4 is 3.7$\times 10^6$~M$_\odot$) and 20 haloes have at least 1 stream within those limits. We find significant halo-to-halo variance between the simulated MW-mass galaxies, suggesting that larges sample of both simulated and observed stream populations is necessary to fully assess the consistency of the MW with predicted stream populations in \LCDM. This finding is consistent with the results of studies of semi-analytic models of large numbers of stream populations around MW-mass hosts \citep{Dropulic:2024}. The effect of halo-to-halo variance, as well as possible trends with host properties, will be discussed in greater detail in \secref{hosts}.

In \figref{peri_apo_l3_full}, there are a small number of stellar streams with $r_{\rm peri} > 150 \kpc$. These are all classified as satellites-of-satellites and have been disrupted by another galaxy than the main host. This effect will be discussed in greater detail in \secref{satsofsats}.

The Auriga stellar streams also span a large range of accretion times and orbital eccentricities, as seen in \figref{peri_apo_l3_full} and \figref{tacc_ecc}. Among the stellar streams, we do not find a strong correlation between accretion time or eccentricity and $f_{\rm bound}$.

Finally, we find no strong preference for either prograde or retrograde orbits among the Auriga stellar stream population as whole. We do however find that some individual simulations have a larger number on either prograde or retrograde orbits, which may be due to the accretion history of the particular host galaxy. This suggests that the fact that the MW streams lie almost exclusively on prograde orbits \citep{Li:2021} may be due to the particular accretion history of the MW rather than a generic outcome of tidal disruption by a MW-like host galaxy.

\subsubsection{Phase-Mixed Systems}
\label{sec:pm_orbs}

Phase-mixed systems in Auriga tend to have early accretion times ($t_{\rm acc} > 8\Gyr$) or highly eccentric orbits ($e > 0.8$), and most have both, as seen in \figref{tacc_ecc}. The right-hand panel of \figref{peri_apo_l3} illustrates the pericentres and apocentres of the phase-mixed systems, which tend to have very small pericentre and apocentre values ($< 10 \kpc$). A relatively small number of systems have larger orbital radii. However, as mentioned above, the orbits of phase-mixed systems are most difficult to define, even in simulations. For all accreted systems we calculate the most recent pericentre and apocentre. Many of the phase-mixed systems have experienced a large number of pericentres and the orbits of their member stars often diverge significantly. The larger values are in many cases therefore not representative of the closest approach distance of the system across its entire history, but of the current median orbit of the member stars. Some of the systems with the largest pericentres have also been identified as satellites-of-satellites and will be discussed in \secref{satsofsats}.

\subsection{Effect of Simulation Resolution}
\label{sec:resolution}

\begin{figure*}
    \includegraphics[width=0.8\textwidth]{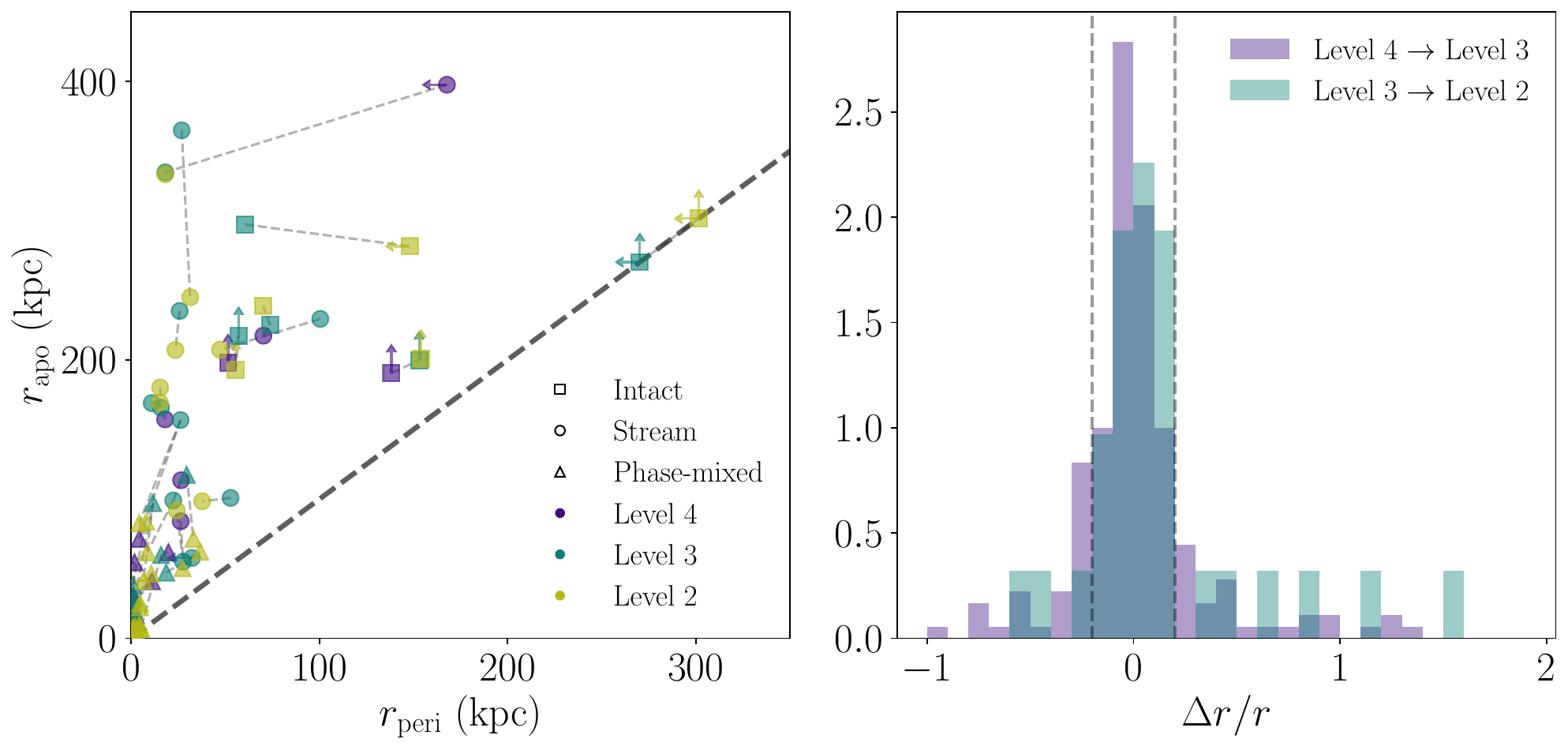}
    \caption{Pericentres and apocentres of satellites of Au-6 across three resolution levels (left). Squares represent intact satellites, circles represent stellar streams, and triangles represent phase-mixed systems. The points are colored by corresponding resolution, from highest resolution level 2 (yellow), to the fiducial resolution level 3 (blue), to the lowest resolution level 4 (purple). Systems that are matches across resolutions are connected by dashed lines. The histograms (right) show the fractional change in pericentres and apocentres of all intact satellites and stellar streams matched between resolution levels. The majority of systems lie within the dashed lines indicating $\pm 20\%$ change. The more significant changes generally correspond to changes in accretion time and thereby in number of pericentres since accretion.}
    \label{fig:peri_apo_res}
\end{figure*}

Satellite galaxy populations in simulations can be impacted by the effects of numerical resolution \citep{Grand:2021}. In order to quantify these effects on the disrupting satellite populations in Auriga, we compare satellites simulated across the three resolutions (level 4, 3, and 2 as described in \secref{sims}).

In particular, we compare the six host galaxies that have been simulated at both level 4 and level 3 resolution, including the single galaxy (Au-6) that has been resimulated at the highest resolution, level 2. For each resimulated galaxy, we match accretion events across resolution by comparing the initial positions of dark matter member particles. This procedure is described in detail in Paper I. 

Paper I also discusses the impact of simulation resolution on disrupting satellite morphology. In short, we find that the morphological classifications are largely converged across resolution. Changes in classification are generally due to minor changes in stellar bound fraction ($f_{\rm bound}$), local velocity dispersion ($\sigma_{\rm 50}$), or accretion time.

Here, we examine the orbits of disrupting satellites across resolution. \figref{peri_apo_res} illustrates the changes in calculated pericentre and apocentre values. The left panel shows systems in Au-6 at all three resolutions, with lines connecting systems that have been matched across resolution. The right panel shows a histogram of the fractional shift in pericentres and apocentres (combined across all resimulated galaxies) from level 4 to 3 and from level 3 to 2. We define $\Delta r/r$ as 
$(r_{\rm low\ res} - r_{\rm high\ res}) / r_{\rm high\ res}.$
The vertical dashed lines indicate a $\pm 20\%$ fractional change. The majority of pericentres and apocentres have a shift of less than $20\%$. For level 4 to 3, the median shift is $-0.5\%$ and 124 of 188 of satellites have shifts smaller than $20\%$. For level 3 to 2, the median shift is $+4\%$ and 22 of 32 satellites have shifts less than $20\%$. We note that the histograms are normalised, and the number of systems in the level 2 histogram is much smaller (32 versus 188), given that only Au-6 has been resimulated at the highest resolution. The median level 3 to 2 shift is positive ($4\%$), possibly suggesting that orbits tend to shift to smaller radii with increasing resolution. However, the sample size is small, and the stellar masses of the Auriga galaxies also tend to increase with increasing resolution \citep{Grand:2017, Grand:2021}, and it is expected that the satellite orbits would move inward as the central mass of the host galaxy increases.  The outliers with $| \Delta r/ r | > 0.5$ are systems that were accreted at slightly different times in the resimulation and therefore have had, for example, one fewer pericentre at one resolution than the other. These slight changes in orbital history are expected between realisations of varying random seed, regardless of change in resolution.

In summary, we find that the orbits of disrupting satellites in Auriga are well-converged across the three resolution levels presented here. Differences in the orbital distributions between the samples at different resolutions are largely due to halo-to-halo scatter and the effects of resimulating galaxies, and not varying systematically due to numerical resolution.

\subsection{Effect of Host Galaxy Properties}
\label{sec:hosts}

The orbits of disrupting satellites are sensitive to the properties of their host galaxies \citep{Garrison-Kimmel:2017, Riley:2019, Dropulic:2024}. Understanding the effect of host properties, including the influence of massive satellites and the dependence on host halo and disc properties is essential for interpreting the MW's own satellite populations in the context of \LCDM.

Here, we consider the 28 haloes simulated at level 4 and study the orbits of their satellite populations as a function of host galaxy and halo properties. As described in \secref{sims}, these haloes were selected based on their mass ($1 - 2 \times 10^{12} \Msun$), and to be isolated at $z=0$ (i.e. none of these galaxies have an Andromeda analog). The selected galaxies otherwise have a range of galaxy properties and accretion histories, which are discussed in more detail elsewhere \citep{Grand:2017, Monachesi:2019, Fattahi:2020, Vera-Casanova:2022}.

\begin{figure*}
    \includegraphics[width=1.0\textwidth]{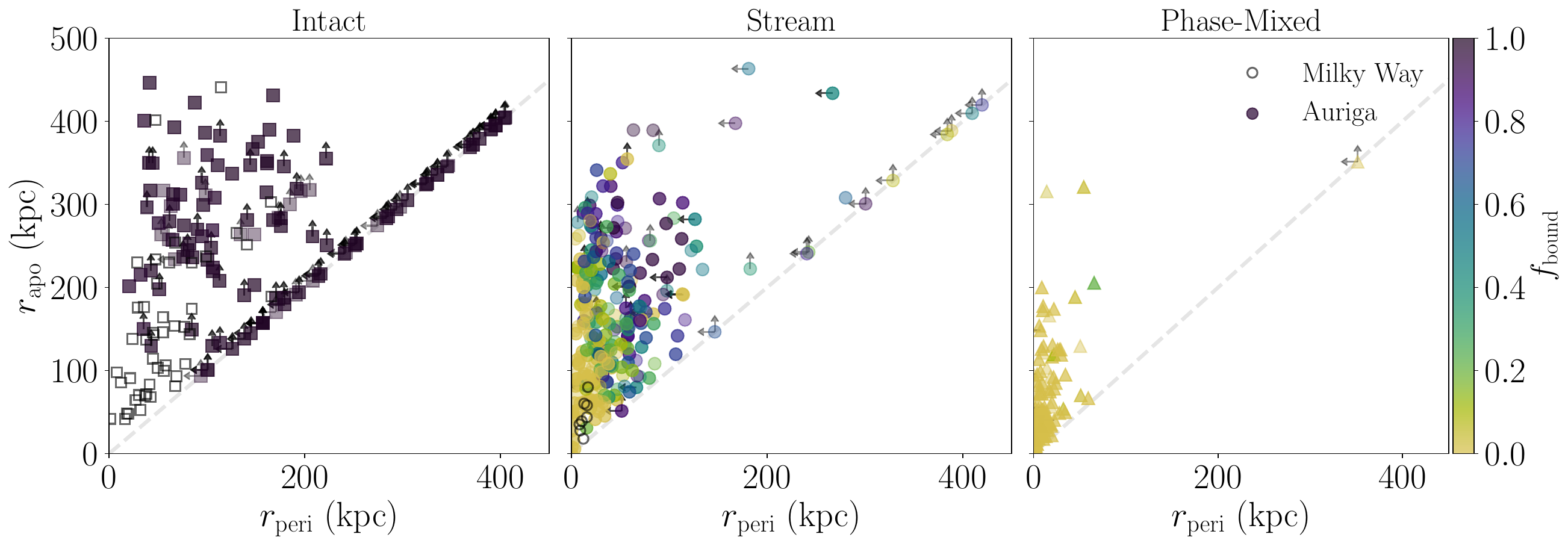}
    \caption{Pericentres and apocentres of intact satellites (left), stellar streams (middle), and phase-mixed systems (right) across the 28 level 4 host galaxies. All points are colored by the stellar mass bound fraction $f_{\rm bound}$. Systems that are classified as satellites-of-satellites are plotted with a higher transparency. As in previous versions of this figure, systems that have not had a pericentre and/or apocentre since accretion are plotted with their current distance as a upper/lower limit on their pericentre/apocentre.}
    \label{fig:peri_apo_l4}
\end{figure*}

\figref{peri_apo_l4} shows the full distribution of pericentres and apocentres of satellites around the 28 level 4 host galaxies, including intact satellites (left panel), stellar streams (middle panel), and phase-mixed systems (right panel). Note that this figure shows the full range of values, unlike in \figref{peri_apo_l3}, which is zoomed in to highlight the region overlapping with MW systems. Across these 28 simulations, there is a large spread in orbital parameters. However, we see that the intact satellites that have had an apocentre since accretion all have $r_{\rm apo} \gtrsim 200$~kpc. This emphasizes the apparent discrepancy with the MW satellite orbits (unfilled squares), many of which have smaller apocentre values. It is unlikely that this discrepancy is entirely due to properties of the MW itself or to halo-to-halo scatter, and supports the conclusion that either the MW satellites are more disrupted than has been so far observed or that a short-coming in our simulations (either numerical or theoretical) is leading to substantial over-disruption of satellite galaxies. Either conclusion would have important implications for studies of satellite galaxies in the context of \LCDM.

A large number of systems across the 28 simulations have not had a pericentre or apocentre since accretion. The majority of these systems are classified as intact satellites, however, there are a significant number of stellar streams and a single phase-mixed system that are newly accreted, currently located at distances of $d > 200 \kpc$, and yet have been tidally disrupted. These systems have been identified as satellites-of-satellites, are indicated by transparent markers, and are discussed in greater detail below.

\subsubsection{Massive Satellites}
\label{sec:satsofsats}

Massive satellites recently accreted on to the the host galaxy can significantly influence the population of surviving and disrupted satellite galaxies. The hierarchical theory of structure formation predicts that galaxies that are accreted on to hosts like the MW should in turn have accreted their own population of lower-mass satellites \citep{Li:2008, Wetzel:2015, Santos-Santos:2021}. In fact, the LMC, the most massive satellite of the MW, is thought to have brought in its own satellites \citep{Kallivayalil:2018, Patel:2020, Nadler:2020, Vasiliev:2024}, and is known to have influenced the orbits, and possibly the disruption rates, of the MW stellar streams \citep[e.g.][]{Gomez:2015, Erkal:2018a, Erkal:2018b, Koposov:2019, Shipp:2021, Vasiliev:2021, Lilleengen:2023, Koposov:2023, Brooks:2024}.

In this work, we classify systems as satellites-of-satellites if they are not the most massive halo in their FOF group (excluding the main host) for three consecutive snapshots (see Paper I for further detail). The satellites are identified in \figref{peri_apo_l4} as the points with a higher transparency. In particular, it is clear that nearly all of the streams that have not had a pericentre and/or apocentre since accretion are classified as satellites-of-satellites and have in fact been disrupted by preprocessing in another environment, not by the tidal field of the main MW-mass host galaxy. The majority of these systems come from a single simulated galaxy, Au-18, which has recently accreted a massive pair of satellites ($M_* = 2.7, 2.3 \times 10^{10} \Msun$).

\begin{figure*}
    \includegraphics[width=0.75\textwidth]{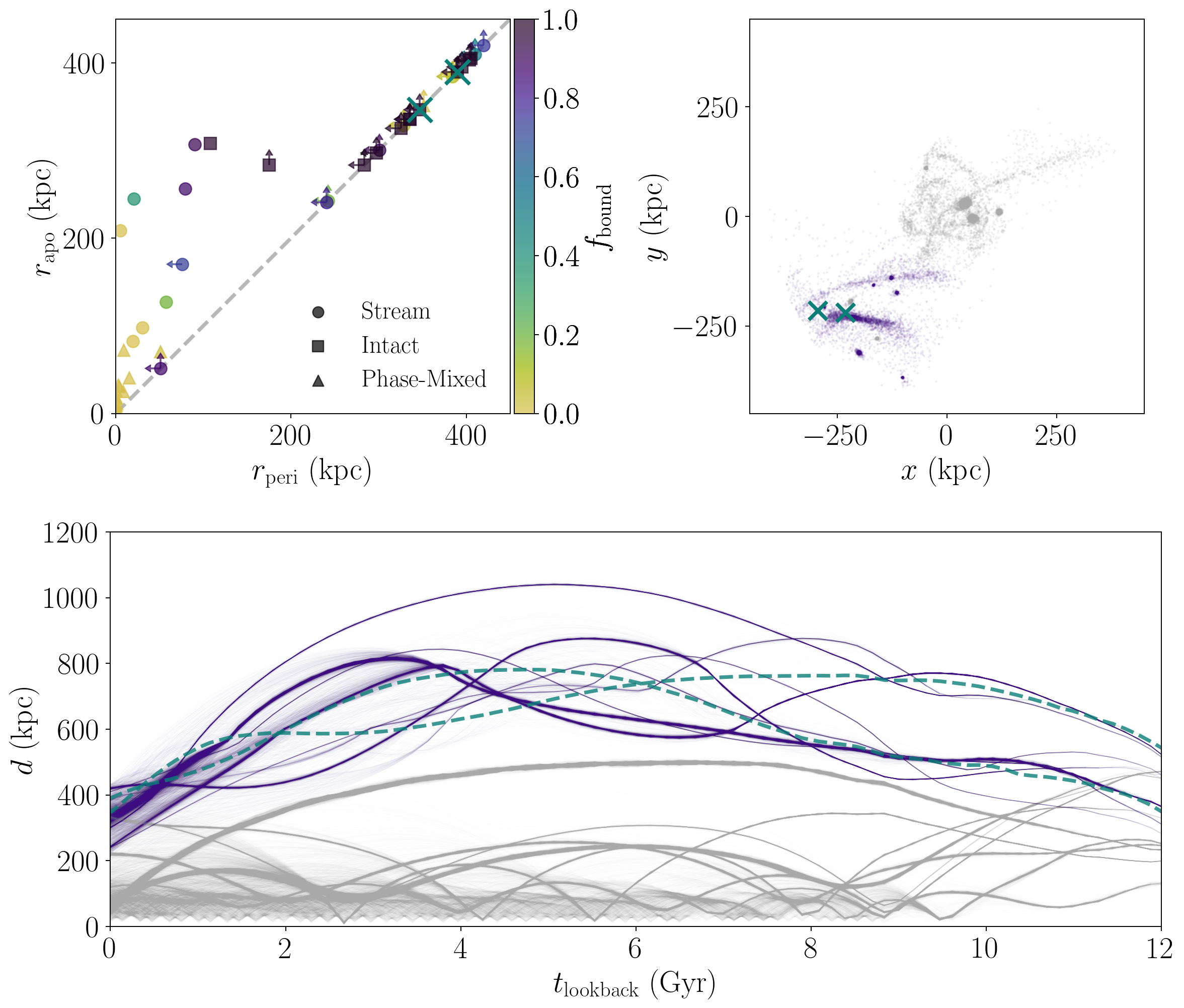}
    \caption{Orbits and positions of disrupting satellites around Au-18, a MW-mass galaxy with a recently accreted, ongoing merger with two massive satellites. The upper-left panel shows the pericentres and apocentres of intact satellites, stellar streams, and phase-mixed systems around Au-18. As in previous versions of this figure, systems that have not yet had a pericentre and/or apocentre after accretion on to the main host are shown with their current distances as upper/lower limits on pericentre/apocentre. The two massive merging satellites are marked as turquoise Xs. The upper right panel shows the $z=0$ spatial distribution of systems classified as stellar streams around Au-18. The systems considered to be satellites of the massive objects are plotted in purple, while all other streams are shown in light gray. The two turquoise Xs indicate the centroids of the massive satellites. The lower panel shows the orbits over time of stellar streams around Au-18. Once again the satellites of the two massive satellites are shown in purple while all other streams are shown in gray. The orbits of the two massive systems are shown as turquoise dashed lines.}
    \label{fig:halo18}
\end{figure*}

These satellites and their associated streams are shown in \figref{halo18}. The top left panel shows the pericentres and apocentres of all the satellites associated with Au-18. There are a large number that are on first infall and have not had a pericentre or apocentre since accretion and are therefore plotted at their current distance along the circular orbit dashed line. The two blue X's indicate the two massive satellites, which are also on first infall into the host galaxy. The spatial distribution of the Au-18 stellar streams is shown in the upper right panel. The blue X's again indicate the location of the two massive satellites. The streams associated with these infalling systems are plotted in purple, while the rest of the Au-18 streams are shown in gray. The lower panel shows the distance over time of all the Au-18 stellar streams, as well as the blue lines indicating the orbits of the two massive satellites. It is clear that several streams are orbiting the massive satellites as the system falls into the main host potential. This explains their high disruption rates despite the lack of close passage to the main host. These streams have instead been disrupted by the massive satellites. The example highlights the importance of considering the effect of massive satellite systems and, more generally, the preprocessing of satellites before infall on to the main host. These effects can influence the overall distribution of pericentres and apocentres and lead to higher disruption rates at large distances from the host galaxy \citep{He:2024}.

\begin{figure}
    \includegraphics[width=0.9\columnwidth]{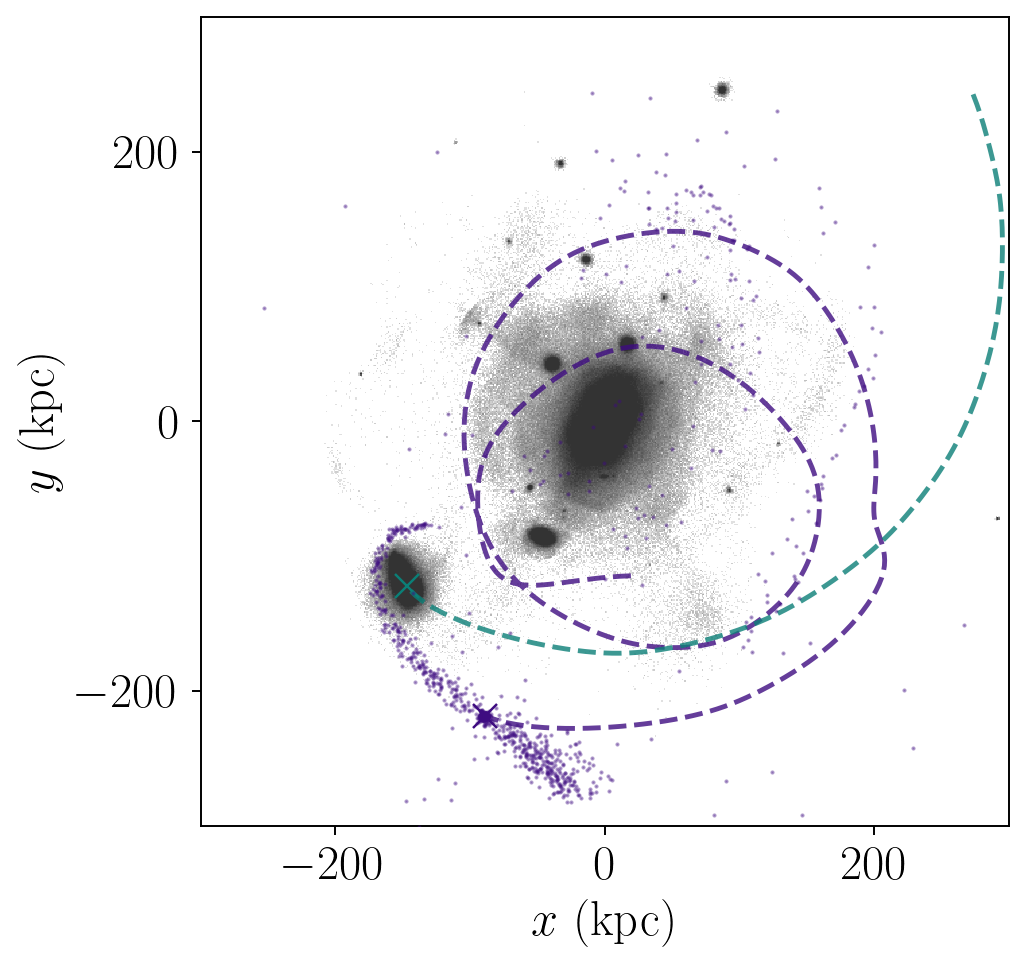}
    \caption{Example of a stellar stream perturbed by a massive satellite. The stream, plotted in purple, has been orbiting the host at a large distance ($r > 100\kpc$) for more than $9 \Gyr$ (past orbit shown as the dashed purple line). However, despite its large distance, it is heavily disrupted ($f_{\rm bound} = 0.36$) due to perturbation by a massive satellite (past orbit shown as turquoise dashed line). The X's indicate the positions at present day of the stream progenitor (purple) and the perturbing massive satellite (turquoise). The background black histogram shows the stars belonging to all of the accreted systems.}
    \label{fig:yoink}
\end{figure}

In addition, massive satellites can have more indirect effects on disrupting satellite populations. Satellites as massive as the LMC are known to cause significant, time-varying distortions to the MW potential \citep{Gomez:2016, Garavito-Camargo:2020} which may have larger-scale effects on the stream population as a whole. For example, \figref{yoink} illustrates a stream that orbited the main host galaxy at a large distance ($r > 100 \kpc$) for more than 9~Gyr with little consequence, and recently lost a substantial fraction of its stellar mass ($f_{\rm bound} = 0.36$) due to a close encounter with a massive satellite. These effects would benefit from further study of cosmological simulations with MW-like merger histories, including a massive LMC-like satellite.

\subsubsection{Halo and Disc Properties}

\begin{figure*}
    \includegraphics[width=1.0\textwidth]{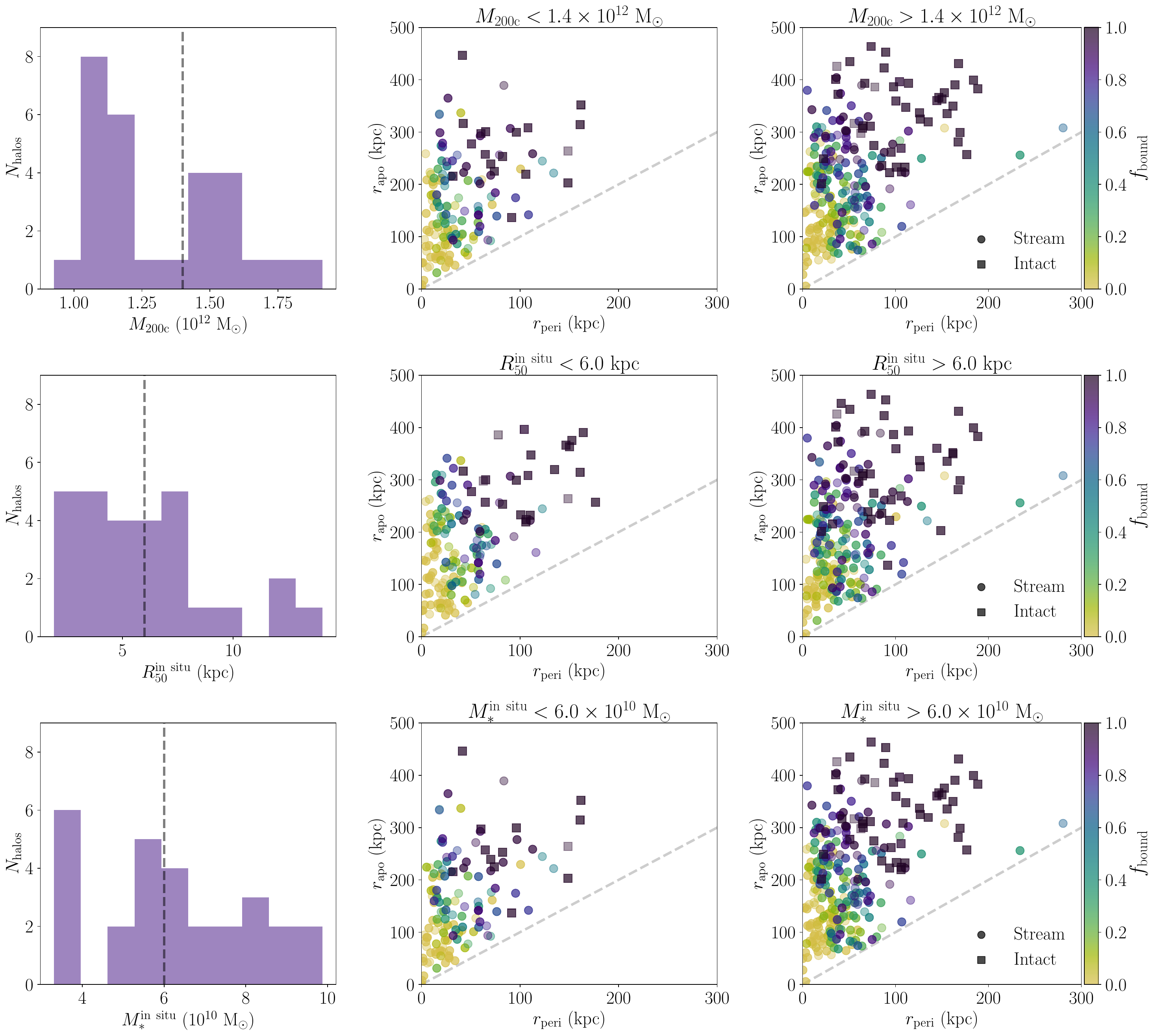}
    \caption{Pericentres and apocentres of intact satellites and stellar streams around 28 level 4 host galaxies binned by host galaxy properties. In the top row, galaxies are binned by host halo mass ($M_{\rm 200c})$. The histogram on the left shows the distribution of host masses across the 28 simulations and the gray dashed line indicates that value on which the sample is split. The panels in the centre and right show the pericentres and apocentres of satellites around galaxies in each mass bin. The middle row shows the same for disc size ($R_{\rm 50}^{\rm in\ situ}$, defined as the spherical radius which encloses $50\%$ of in situ stellar mass), and the bottom row shows the distribution across host in situ stellar mass ($M_{\rm *}^{\rm in\ situ}$). 
    }
    \label{fig:host_props}
\end{figure*}

The properties of the host galaxy itself can also have a significant effect on disrupting satellite populations. In particular, the host mass is correlated with the number, mass function, and radial distribution of infalling satellites, and the disc contributes significantly to satellite disruption \citep{Garrison-Kimmel:2017, Errani:2017, Nadler:2018, Kelley:2019, Wang:2024}. These effects propagate to the orbital distributions of satellites at all stages of tidal disruption.

In \figref{host_props}, we examine how the distributions of pericentres and apocentres of intact satellites and stellar streams depend on properties of the host halo and central disc. Once again, we consider the 28 level 4 simulations. Here we show the distributions binned by host halo mass ($M_\text{200c}$, top row), host galaxy disc size ($R_{\rm 50}^{\rm in\ situ}$, defined as the spherical radius that encloses $50\%$ of in situ stellar mass, middle row), and in situ stellar mass ($M_{\rm *}^{\rm in\ situ}$, a proxy for disc mass, bottom row). These values for each simulation are compiled in Table 1 of Paper I.

We find suggestions of dependence on the stream and intact satellite pericentre and apocentre distributions on each of these parameters. First, satellites at all stages of disruption seem to shift to larger orbits with increasing $M_{\rm 200c}$, which is expected as the size of the host galaxy potential grows. Although expected, this is an important point, given the significant uncertainty on the halo mass of the MW \citep{Callingham:2019, Wang:2020}. Understanding discrepancies between predicted and observed orbits of disrupting satellites requires accounting for these uncertainties on the properties of the MW itself.

Second, satellites at smaller orbital radii seem to be less disrupted (larger $f_{\rm bound}$) with increasing $R_{\rm 50}^{\rm in\ situ}$. Among this sample of galaxies, the disc size is not highly correlated with either disc mass or halo mass. This means that increasing $R_{\rm 50}^{\rm in\ situ}$ tends to decrease the density of the disc, which may allow satellites to retain more of their member stars while orbiting deeper in the galactic potential. Finally, $M_{\rm *}^{\rm in\ situ}$ tends to be highly correlated with the halo mass. It is therefore unsurprising that we see a similar increase in orbital radii with increasing in situ stellar mass as with increasing $M_{\rm 200c}$.

We also considered other host parameters, including accreted stellar mass, in situ star formation time-scale (as a proxy for disc formation time), the stellar mass of the largest satellite, and the recent accretion rate of the host galaxy.\footnote{We define the recent mass accretion rate of the host between $z=0-0.5$ as $\Gamma \equiv \Delta \log(M_{\rm 200c}) / \Delta \log(a)$, from \citet{Deimer:2014}.} When normalizing orbits by viral radius $R_{\rm 200c}$ in order to remove the effects of correlation with increasing halo mass and size, we find no clear trends with any of these properties.

Further studies are called for to understand the significance of the trends highlighted in \figref{host_props} and to investigate trends with other host galaxy and halo properties. \citet{Dropulic:2024} study the effect of disrupting satellite orbits on host properties in semi-analytic models and identify similar trends with halo mass, disc mass, and disc density. Studies of the effect of host properties on disruption satellite populations are important while we are limited to a single observed galaxy (the MW). They will also be essential as we extend these analyses to external galaxies with upcoming observations from facilities such as Euclid \citep{Racca:2016}, the Vera C. Rubin Observatory \citep{LSST:2019}, the Nancy Grace Roman Space Telescope \citep{Spergel:2013}, as well as from the ESA-selected ARRAKIHS\footnote{\href{https://www.cosmos.esa.int/documents/7423467/7423486/ESA-F2-ARRAKIHS-Phase-2-PUBLIC-v0.9.2.pdf/61b363d7-2a06-1196-5c40-c85aa90c2113?t=1667557422996}{ARRAKIHS Phase 2 Proposal (public, pdf)}} mission.

\section{Discussion and Conclusion}
\label{sec:discussion}

In summary, we find that satellites around MW-mass galaxies in the Auriga simulations are highly disrupted, with the majority of `surviving' satellites being classified as stellar streams (i.e. they have extended, coherent tidal tails). We also find that the subset of the satellite population that remains fully intact to $z=0$ 
are recently accreted ($n_{\rm peri} \lesssim 1$) and have large pericentres ($r_{\rm peri} \gtrsim 50 \kpc$) and apocentres ($r_{\rm apo} \gtrsim 200 \kpc$). The disrupted satellites, however, span a much larger range of accretion times, pericentres, and apocentres. There is a strong apparent discrepancy in both the number and the orbits of the fully intact satellites in the Auriga simulations and those observed around the MW. There are more known MW satellites (in the resolved mass range) than the number of \textit{intact} satellites in any Auriga simulation, and the orbits of the surviving MW satellites overlap significantly with both the intact and disrupting satellite population in Auriga. This apparent discrepancy may be resolved if the MW satellites are in fact \textit{more highly disrupted} than has previously been observed. This higher disruption rate may also help to explain recent observations of extra-tidal stars around MW satellite galaxies, including those with large pericentric distances \citep[e.g.][]{Jensen:2024}.

In the future, we will generate mock-observed catalogs of resolved stellar populations in the simulated satellites and compare to detectable satellites and tidal tails in the MW in order to determine whether this predicted high disruption rate of MW satellites is in fact consistent with current observations.

Another key finding is that many of the Auriga simulations have stellar streams on orbits consistent with the MW dwarf galaxy streams ($r_{\rm peri} \lesssim 20 \kpc, r_{\rm apo} \lesssim 50 \kpc)$. This is different from the findings of \citet{Shipp:2023} using FIRE-2 simulations. However, among the larger sample of 28 Auriga simulations, we find significant halo-to-halo variance \citep[a conclusion also supported by semi-analytic models in][]{Dropulic:2024}, with some Auriga simulations more similar to the MW distribution, and some more similar to FIRE-2. Therefore, given the relatively small sample size (28 galaxies in Auriga, 13 in FIRE-2), it is difficult to tell whether any observed differences in orbital properties are due to the effect of halo-to-halo variance or in fact due to model-based differences. It is possible that the 13 FIRE-2 galaxies studied in \citet{Shipp:2023} just happened to be systems with streams orbiting at larger distances from the centre of the host galaxy. 

If these 13 galaxies are in fact representative of all FIRE-2 galaxies, the lack of streams at small pericentres and apocentres could be due to a higher rate of phase-mixing in the inner galaxy (e.g. due to more time-dependence in the central potential) than in the Auriga simulations. It does not appear that the difference lies in the initial satellite disruption rates, because the orbits of the intact satellite population appear to be consistent between the simulations -- there is only a difference in whether certain disrupting satellites in the inner galaxy are classified as coherent stellar streams or phase-mixed systems. Interestingly, \citet{Santistevan:2023} found that the orbits of surviving satellites in FIRE-2 tend to increase over time, possibly due to time-evolution of the host galaxy potential. This effect could also cause the streams to orbit at larger radii at $z=0$. Importantly, both Auriga and FIRE form massive central discs, so purely the presence of a host galaxy is likely not driving the differences between these simulations and the observations.

On the other hand, we find a correlation between stream orbits and host halo mass \citep[also supported by][]{Dropulic:2024}, with less massive host galaxies having streams with smaller pericentres and apocentres. If the MW's halo mass is less than $10^{12}\Msun$, as suggested by some recent measurements \citep[e.g.][though see \citealt{Oman:2024}]{Vasiliev:2021, Koposov:2023, Ou:2024rotcurve}, we would expect the streams to be orbiting at smaller distances than predicted by either the FIRE-2 or Auriga simulations, which include host galaxies with masses of $1-2 \times 10^{12} \Msun$. These results call for studies of large numbers of stream populations with a range of host galaxy properties in both cosmological simulations and in semi-analytic models.

\subsection{Simulation Effects}
\label{sec:sim_effects}

Overall, the Auriga and FIRE-2 simulations predict very similar orbital distributions of disrupting satellites. In both simulations, the majority of satellites are disrupted and even the `surviving' satellites have extended tidal tails and would be classified as stellar streams given perfect observations. There may be a difference in the orbits or phase-mixing rates of streams at small pericentres and apocentres, but as discussed above, this may be due solely to small sample sizes and the effect of halo-to-halo variance. If the disruption rates in these simulations are correct, the lack of observed tidal tails around Milky Way satellites could be due to their low surface brightness, a scenario verified for FIRE-2 \citep{Shipp:2023} but not yet for Auriga. The agreement between these distinct simulations strengthens the confidence in each of their predictions. This agreement is especially interesting given the differences between these sets of simulations. 

One notable difference lies in the treatment of feedback processes and baryonic physics. The FIRE-2 simulations include detailed models of supernovae and stellar winds, attempting to directly capture these feedback processes \citep[see the complete description in][]{Hopkins:2018, Wetzel:2023}. In contrast, Auriga adopts an effective feedback model, where isotropic winds are launched stochastically from supernova sites and are initially decoupled from the gas hydrodynamically until the wind particle enters a predetermined threshold density \citep[at scales larger than star forming regions; see][]{Grand:2017}. Additionally, the two simulations differ in their treatment of the interstellar medium. Auriga employs an effective model for the multiphase gas down to $10^4$~K, while FIRE-2 aims to resolve individual gas phases and include low-temperature cooling down to $10$~K. Despite these different methods, the gas mass resolution remains comparable between FIRE-2 and Auriga level 3 \citep{Wetzel:2016, Grand:2017}. Finally, the two simulations use different gravitational solvers and softening parameters, including variations in the softening lengths \citep{Springel:2010, Hopkins:2015, Wetzel:2016, Grand:2017, Grand:2024}. These modelling choices may lead to discrepancies in the structure and time evolution of both satellites and host galaxies, which could affect disruption rates. However, given the subtleties and complexities of these models, the specific impact of these differences on satellite dynamics requires further targeted investigation. Purely numerical effects could also be tested through comparison projects like AGORA, which examine the influence of simulation techniques on satellite evolution \citep{Augora2014, AugoraSatellite2024}.

On the other hand, both the Auriga and FIRE-2 simulations are susceptible to certain numerical issues. These challenges are particularly evident in the number of particles used to sample the dark matter and stellar distribution functions of the satellites, as well as in potential discrepancies with particle mass ratios. Tailored $N$-body simulations \citep{vandenBosch:2018, Errani:2020, green2021tidal, Errani:2021} have demonstrated lower subhalo disruption rates, suggesting that satellites may disrupt artificially when the particle count or force resolution is insufficient. However, these controlled experiments do not fully replicate the complexities of a cosmological environment, such as satellite preprocessing or the effects of a central massive disc and time-evolving potential, which are crucial for accurately modeling satellite evolution \citep{Garrison-Kimmel:2017, Kelley:2019, Wang:2024, He:2024}\footnote{For further discussions of numerical tests of these effects in a cosmological context, see \citet{Grand:2021, barry2023dark, donlon2024debris}.}.
Spurious heating due to unequal dark matter and star particle masses or insufficient particle number could also cause the satellites to disrupt more quickly than they should \citep[see][]{Ludlow:2019, Ludlow:2020, Ludlow:2021, Ludlow:2023}. However, this effect has not been thoroughly tested for satellite systems (as opposed to central galaxies), and further study is needed, likely involving high-resolution simulations and convergence tests across different simulation suites. 

\subsection{Implications for Milky Way Satellites}
\label{sec:mw_sats}

If the disruption rates predicted by cosmological simulations like Auriga are indeed correct, then the MW satellites are likely much more disrupted than we have yet to observe. This is not necessarily inconsistent with current observations, given that the tidal tails tend to be quite diffuse and low surface brightness, and in many cases the surviving bound components of the galaxies remain relatively undisturbed. 

Future observations may reveal evidence of tidal disruption around existing satellites. In particular, deep photometric observations with surveys like the Rubin Observatory Legacy Survey of Space and Time \citep[LSST;][]{LSST:2009} are well-suited to reveal low surface brightness stellar density features like stellar streams \citep{LSST:2019}. Narrow band photometric surveys targeting metallicity-sensitive lines such as CaHK \citep[e.g.][]{Starkenburg:2017} are also powerful tools in selecting the metal-poor member stars of known satellites across larger areas. Proper motions from \Gaia have been used to select member stars out to larger distances around known satellites, as in \citet{Jensen:2024}, and future \Gaia data releases will provide even higher precision proper motion measurements that may enable improved selection of member stars beyond the tidal radii of these satellites. Finally, wide-field spectroscopic surveys such as the Dark Energy Spectroscopic Instrument \citep[DESI; ][]{Cooper:2023, Koposov:2024}, 4MOST \citep{4MOST:2019}, and WEAVE \citep{WEAVE:2016} will also facilitate the identification of members across a wider area than more targeted spectroscopic observations. Furthermore, when taken together, these surveys will provide full kinematic measurements of a large sample of member stars of these satellites, enabling dynamical models that may be used to reveal the disruption history of the MW satellite population. Future work will consider the disrupting satellites in Auriga in the context of the MW and make predictions for observational measurements that may be used to further test the predictions of these simulations in comparison to the MW satellite population. Finally, future observations with LSST, Euclid \citep{Racca:2016}, the Nancy Grace Roman Space telescope \citep{Spergel:2013}, and the ESA-selected ARRAKHHIS mission will also reveal populations of satellites and stellar streams around MW-mass hosts external to our own galaxy, thereby providing a larger sample size with which to compare our simulations.

High stellar disruption rates also have implications for the dark matter components of the MW satellite galaxies. The majority of the dark matter is stripped before these systems begin losing stars, so satellites that have experienced the level of stellar disruption that we see in Auriga would have very low-mass dark matter components. This could bias dwarf galaxy mass modeling measurements and dark matter indirect detection constraints, though constraints based on comparisons to cosmological simulations would already incorporate these effects \citep[e.g.][]{Wang:2022, Vienneau:2024}. The detailed dark matter distributions of the Auriga satellites and the resulting effect on dark matter constraints using MW satellites will be examined in future work.

As discussed above, there are many assumptions and approximations that go into these complex simulations that could potentially lead to artificially high tidal disruption rates of the Auriga satellite population. If instead, the Auriga and FIRE-2 simulations are over-disrupting satellites, the results presented here still have important implications for studies of the MW satellite population in the context of predictions of \LCDM and galaxy formation models. Many suites of hydrodynamic cosmological simulations, including Auriga, do an excellent job of reproducing many properties of the surviving satellite population as observed around the MW. However, if these satellites are in fact over-disrupted, then there must be significant remaining discrepancies, including in the total mass function of satellites (the total masses of Auriga satellites are higher than their present day masses given the significant tidal mass loss), the structures of satellite galaxies (density profiles of satellites affect disruption rates), and the orbits of satellites and their radial distributions within the MW. These discrepancies would indicate significant remaining gaps in our understanding of small-scale galaxy formation in the context of \LCDM. Truly understanding whether our simulations of satellite galaxy populations in \LCDM are able to reproduce MW observations will require a more thorough understanding of disruption rates of satellite galaxies in both simulations and in observations.

\section*{Acknowledgements}
The authors would like to thank Denis Erkal, Phil Mansfield, Robyn Sanderson, and Andrew Wetzel for valuable conversations. This research was supported in part by grant NSF PHY-2309135 to the Kavli Institute for Theoretical Physics (KITP). This project was developed in part at the Streams24 meeting hosted at Durham University. NS was supported by an NSF Astronomy and Astrophysics Postdoctoral Fellowship under award AST-2303841. AHR is supported by a Research Fellowship from the Royal Commission for the Exhibition of 1851 and by STFC through grant ST/T000244/1. RB is supported by the UZH Postdoc Grant, grant no. FK-23116 and the SNSF through the Ambizione Grant PZ00P2\_223532. LN is supported by the Sloan Fellowship, the NSF CAREER award 2337864, NSF award 2307788, and NSF award PHY-2019786 (The NSF AI Institute for Artificial Intelligence and Fundamental Interactions, http://iaifi.org/). FF is supported by a UKRI Future Leaders Fellowship (grant no. MR/X033740/1). FAG acknowledges funding from the Max Planck Society through a “PartnerGroup” grant. FAG acknowledges support from ANID FONDECYT Regular 1211370, the ANID Basal Project FB210003 and the HORIZON-MSCA-2021-SE-01 Research and innovation programme under the Marie Sklodowska-Curie grant agreement number 101086388. RJJG acknowledges financial support from an STFC Ernest Rutherford Fellowship (ST/W003643/1). FM acknowledges funding from the European Union - NextGenerationEU under the HPC project `National Centre for HPC, Big Data and Quantum Computing' (PNRR - M4C2 - I1.4 - CN00000013 – CUP J33C22001170001). 

This research used resources of the Argonne Leadership Computing Facility, a U.S. Department of Energy (DOE) Office of Science user facility at Argonne National Laboratory and is based on research supported by the U.S. DOE Office of Science-Advanced Scientific Computing Research Program, under Contract No. DE-AC02-06CH11357. This work used the Freya computer cluster at the Max Planck Institute for Astrophysics. This work used the DiRAC@Durham facility managed by the Institute for Computational Cosmology on behalf of the STFC DiRAC HPC Facility (www.dirac.ac.uk). The equipment was funded by BEIS capital funding via STFC capital grants ST/K00042X/1, ST/P002293/1, ST/R002371/1 and ST/S002502/1, Durham University and STFC operations grant ST/R000832/1.
DiRAC is part of the National e-Infrastructure.

\section*{Software}
This research made use of the Python programming language, along with many community-developed or maintained software packages including:
\begin{itemize}
    \item AGAMA \citep{agama}
    \item Astropy \citep{Astropy:2013, Astropy:2018, Astropy:2022}
    \item CMasher \citep{cmasher, cmasher1.8.0}
    \item Cython \citep{cython}
    \item h5py \citep{h5py, h5py3.7.0}
    \item Jupyter \citep{ipython, jupyter}
    \item Matplotlib \citep{matplotlib}
    \item NumPy \citep{numpy}
    \item Pandas \citep{pandas, pandas1.5.0}
    \item Scikit-learn \citep{scikit-learn, scikit-learn-api, scikit-learn1.1.2}
    \item SciPy \citep{scipy, scipy1.9.1}
\end{itemize}
We also thank the maintainers of the arepo-snap-util package, as well as Victor Forouhar Moreno for fusing the \citet{Richings:2020} framework with AGAMA.
Parts of the results in this work make use of the colormaps in the CMasher package.
Software citation information aggregated using \href{https://www.tomwagg.com/software-citation-station/}{The Software Citation Station} \citep{software-citation-station-paper, software-citation-station-zenodo}.

\section*{Data Availability}

Halo catalogs, merger trees, and particle data (\secref{sims}) for Auriga levels 3 and 4 are publicly available \citep[detailed in the Auriga project data release;][]{Grand:2024} to download via the Globus platform\footnote{https://wwwmpa.mpa-garching.mpg.de/auriga/data}. Auriga level 2 data products will be shared upon reasonable request. The orbits and properties of the disrupting satellite galaxies characterised in this article (\secref{results}) are available in Appendix~\ref{app:datatables} and on \href{https://wwwmpa.mpa-garching.mpg.de/auriga/dataspecs.html#Highleveldata}{the Auriga webpage}.


\bibliographystyle{mnras}
\bibliography{main, software-alex}

\begin{thebibliography}{}
\makeatletter
\relax
\def\mn@urlcharsother{\let\do\@makeother \do\$\do\&\do\#\do\^\do\_\do\%\do\~}
\def\mn@doi{\begingroup\mn@urlcharsother \@ifnextchar [ {\mn@doi@}
  {\mn@doi@[]}}
\def\mn@doi@[#1]#2{\def\@tempa{#1}\ifx\@tempa\@empty \href
  {http://dx.doi.org/#2} {doi:#2}\else \href {http://dx.doi.org/#2} {#1}\fi
  \endgroup}
\def\mn@eprint#1#2{\mn@eprint@#1:#2::\@nil}
\def\mn@eprint@arXiv#1{\href {http://arxiv.org/abs/#1} {{\tt arXiv:#1}}}
\def\mn@eprint@dblp#1{\href {http://dblp.uni-trier.de/rec/bibtex/#1.xml}
  {dblp:#1}}
\def\mn@eprint@#1:#2:#3:#4\@nil{\def\@tempa {#1}\def\@tempb {#2}\def\@tempc
  {#3}\ifx \@tempc \@empty \let \@tempc \@tempb \let \@tempb \@tempa \fi \ifx
  \@tempb \@empty \def\@tempb {arXiv}\fi \@ifundefined
  {mn@eprint@\@tempb}{\@tempb:\@tempc}{\expandafter \expandafter \csname
  mn@eprint@\@tempb\endcsname \expandafter{\@tempc}}}

\bibitem[\protect\citeauthoryear{Arora, Sanderson, Panithanpaisal, Cunningham,
  Wetzel  \& Garavito-Camargo}{Arora et~al.}{2022}]{Arora:2022}
Arora A.,  Sanderson R.~E.,  Panithanpaisal N.,  Cunningham E.~C.,  Wetzel A.,
   Garavito-Camargo N.,  2022, \mn@doi [The Astrophysical Journal]
  {10.3847/1538-4357/ac93fb}, 939, 2

\bibitem[\protect\citeauthoryear{{Arora} et~al.,}{{Arora}
  et~al.}{2024}]{Arora:2024}
{Arora} A.,  et~al., 2024, \mn@doi [arXiv e-prints]
  {10.48550/arXiv.2407.12932}, \href
  {https://ui.adsabs.harvard.edu/abs/2024arXiv240712932A} {p. arXiv:2407.12932}

\bibitem[\protect\citeauthoryear{{Astropy Collaboration} et~al.,}{{Astropy
  Collaboration} et~al.}{2013}]{Astropy:2013}
{Astropy Collaboration} et~al., 2013, \mn@doi [\aap]
  {10.1051/0004-6361/201322068}, \href
  {https://ui.adsabs.harvard.edu/\#abs/2013A&A...558A..33A} {558, A33}

\bibitem[\protect\citeauthoryear{{Astropy Collaboration} et~al.,}{{Astropy
  Collaboration} et~al.}{2018}]{Astropy:2018}
{Astropy Collaboration} et~al., 2018, \mn@doi [\aj] {10.3847/1538-3881/aabc4f},
  \href {https://ui.adsabs.harvard.edu/\#abs/2018AJ....156..123A} {156, 123}

\bibitem[\protect\citeauthoryear{{Astropy Collaboration} et~al.,}{{Astropy
  Collaboration} et~al.}{2022}]{Astropy:2022}
{Astropy Collaboration} et~al., 2022, \mn@doi [\apj]
  {10.3847/1538-4357/ac7c74}, \href
  {https://ui.adsabs.harvard.edu/abs/2022ApJ...935..167A} {935, 167}

\bibitem[\protect\citeauthoryear{Barry, Wetzel, Chapman, Samuel, Sanderson  \&
  Arora}{Barry et~al.}{2023}]{barry2023dark}
Barry M.,  Wetzel A.,  Chapman S.,  Samuel J.,  Sanderson R.,   Arora A.,
  2023, \mnras, 523, 428

\bibitem[\protect\citeauthoryear{Behnel, Bradshaw, Citro, Dalcin, Seljebotn  \&
  Smith}{Behnel et~al.}{2011}]{cython}
Behnel S.,  Bradshaw R.,  Citro C.,  Dalcin L.,  Seljebotn D.~S.,   Smith K.,
  2011, \mn@doi [Computing in Science Engineering] {10.1109/MCSE.2010.118}, 13,
  31

\bibitem[\protect\citeauthoryear{{Belokurov} et~al.,}{{Belokurov}
  et~al.}{2006}]{Belokurov:2006}
{Belokurov} V.,  et~al., 2006, \mn@doi [\apjl] {10.1086/504797}, \href
  {http://adsabs.harvard.edu/abs/2006ApJ...642L.137B} {642, L137}

\bibitem[\protect\citeauthoryear{{Belokurov}, {Erkal}, {Evans}, {Koposov}  \&
  {Deason}}{{Belokurov} et~al.}{2018}]{Belokurov:2018}
{Belokurov} V.,  {Erkal} D.,  {Evans} N.~W.,  {Koposov} S.~E.,   {Deason}
  A.~J.,  2018, \mn@doi [\mnras] {10.1093/mnras/sty982}, \href
  {https://ui.adsabs.harvard.edu/abs/2018MNRAS.478..611B} {478, 611}

\bibitem[\protect\citeauthoryear{{Bernard} et~al.,}{{Bernard}
  et~al.}{2016}]{Bernard:2016}
{Bernard} E.~J.,  et~al., 2016, \mn@doi [\mnras] {10.1093/mnras/stw2134}, \href
  {http://adsabs.harvard.edu/abs/2016MNRAS.463.1759B} {463, 1759}

\bibitem[\protect\citeauthoryear{{Bonaca} \& {Price-Whelan}}{{Bonaca} \&
  {Price-Whelan}}{2024}]{BonacaPriceWhelan:2024}
{Bonaca} A.,  {Price-Whelan} A.~M.,  2024, \mn@doi [arXiv e-prints]
  {10.48550/arXiv.2405.19410}, \href
  {https://ui.adsabs.harvard.edu/abs/2024arXiv240519410B} {p. arXiv:2405.19410}

\bibitem[\protect\citeauthoryear{{Bose}, {Deason}, {Belokurov}  \&
  {Frenk}}{{Bose} et~al.}{2020}]{Bose:2020}
{Bose} S.,  {Deason} A.~J.,  {Belokurov} V.,   {Frenk} C.~S.,  2020, \mn@doi
  [\mnras] {10.1093/mnras/staa1199}, \href
  {https://ui.adsabs.harvard.edu/abs/2020MNRAS.495..743B} {495, 743}

\bibitem[\protect\citeauthoryear{{Brooks} \& {Zolotov}}{{Brooks} \&
  {Zolotov}}{2014}]{Brooks:2014}
{Brooks} A.~M.,  {Zolotov} A.,  2014, \mn@doi [\apj]
  {10.1088/0004-637X/786/2/87}, \href
  {https://ui.adsabs.harvard.edu/abs/2014ApJ...786...87B} {786, 87}

\bibitem[\protect\citeauthoryear{{Brooks}, {Garavito-Camargo}, {Johnston},
  {Price-Whelan}, {Sanders}  \& {Lilleengen}}{{Brooks}
  et~al.}{2024}]{Brooks:2024}
{Brooks} R. A.~N.,  {Garavito-Camargo} N.,  {Johnston} K.~V.,  {Price-Whelan}
  A.~M.,  {Sanders} J.~L.,   {Lilleengen} S.,  2024, \mn@doi [arXiv e-prints]
  {10.48550/arXiv.2410.02574}, \href
  {https://ui.adsabs.harvard.edu/abs/2024arXiv241002574B} {p. arXiv:2410.02574}

\bibitem[\protect\citeauthoryear{Buitinck et~al.,}{Buitinck
  et~al.}{2013}]{scikit-learn-api}
Buitinck L.,  et~al., 2013, in ECML PKDD Workshop: Languages for Data Mining
  and Machine Learning. pp 108--122

\bibitem[\protect\citeauthoryear{{Bullock} \& {Boylan-Kolchin}}{{Bullock} \&
  {Boylan-Kolchin}}{2017}]{Bullock:2017}
{Bullock} J.~S.,  {Boylan-Kolchin} M.,  2017, \mn@doi [\araa]
  {10.1146/annurev-astro-091916-055313}, \href
  {http://adsabs.harvard.edu/abs/2017ARA%26A..55..343B} {55, 343}

\bibitem[\protect\citeauthoryear{{Bullock} \& {Johnston}}{{Bullock} \&
  {Johnston}}{2005}]{Bullock:2005}
{Bullock} J.~S.,  {Johnston} K.~V.,  2005, \mn@doi [\apj] {10.1086/497422},
  \href {http://adsabs.harvard.edu/abs/2005ApJ...635..931B} {635, 931}

\bibitem[\protect\citeauthoryear{{Callingham} et~al.,}{{Callingham}
  et~al.}{2019}]{Callingham:2019}
{Callingham} T.~M.,  et~al., 2019, \mn@doi [\mnras] {10.1093/mnras/stz365},
  \href {https://ui.adsabs.harvard.edu/abs/2019MNRAS.484.5453C} {484, 5453}

\bibitem[\protect\citeauthoryear{{Carlin} \& {Sand}}{{Carlin} \&
  {Sand}}{2018}]{Carlin:2018}
{Carlin} J.~L.,  {Sand} D.~J.,  2018, \mn@doi [\apj]
  {10.3847/1538-4357/aad8c1}, \href
  {https://ui.adsabs.harvard.edu/abs/2018ApJ...865....7C} {865, 7}

\bibitem[\protect\citeauthoryear{{Chiti} et~al.,}{{Chiti}
  et~al.}{2021}]{Chiti:2021}
{Chiti} A.,  et~al., 2021, \mn@doi [Nature Astronomy]
  {10.1038/s41550-020-01285-w}, \href
  {https://ui.adsabs.harvard.edu/abs/2021NatAs...5..392C} {5, 392}

\bibitem[\protect\citeauthoryear{Collette}{Collette}{2013}]{h5py}
Collette A.,  2013, Python and HDF5.
O'Reilly

\bibitem[\protect\citeauthoryear{Collette et~al.,}{Collette
  et~al.}{2022}]{h5py3.7.0}
Collette A.,  et~al., 2022, h5py/h5py: 3.7.0, \mn@doi{10.5281/zenodo.6575970},
  \url {https://doi.org/10.5281/zenodo.6575970}

\bibitem[\protect\citeauthoryear{{Cooper} et~al.,}{{Cooper}
  et~al.}{2023}]{Cooper:2023}
{Cooper} A.~P.,  et~al., 2023, \mn@doi [\apj] {10.3847/1538-4357/acb3c0}, \href
  {https://ui.adsabs.harvard.edu/abs/2023ApJ...947...37C} {947, 37}

\bibitem[\protect\citeauthoryear{{DES Collaboration}}{{DES
  Collaboration}}{2016}]{DES:2016}
{DES Collaboration} 2016, \mn@doi [\mnras] {10.1093/mnras/stw641}, \href
  {http://adsabs.harvard.edu/abs/2016MNRAS.460.1270D} {460, 1270}

\bibitem[\protect\citeauthoryear{{D'Souza} \& {Bell}}{{D'Souza} \&
  {Bell}}{2022}]{D'Souza:2022}
{D'Souza} R.,  {Bell} E.~F.,  2022, \mn@doi [\mnras] {10.1093/mnras/stac404},
  \href {https://ui.adsabs.harvard.edu/abs/2022MNRAS.512..739D} {512, 739}

\bibitem[\protect\citeauthoryear{{Dalton}}{{Dalton}}{2016}]{WEAVE:2016}
{Dalton} G.,  2016, in {Skillen} I.,  {Balcells} M.,   {Trager} S.,  eds,
  Astronomical Society of the Pacific Conference Series Vol. 507, Multi-Object
  Spectroscopy in the Next Decade: Big Questions, Large Surveys, and Wide
  Fields. p.~97

\bibitem[\protect\citeauthoryear{{Davis}, {Efstathiou}, {Frenk}  \&
  {White}}{{Davis} et~al.}{1985}]{Davis:1985}
{Davis} M.,  {Efstathiou} G.,  {Frenk} C.~S.,   {White} S.~D.~M.,  1985,
  \mn@doi [\apj] {10.1086/163168}, \href
  {https://ui.adsabs.harvard.edu/abs/1985ApJ...292..371D} {292, 371}

\bibitem[\protect\citeauthoryear{{Deason}, {Belokurov}, {Koposov}, {G{\'o}mez},
  {Grand}, {Marinacci}  \& {Pakmor}}{{Deason} et~al.}{2017}]{Deason:2017}
{Deason} A.~J.,  {Belokurov} V.,  {Koposov} S.~E.,  {G{\'o}mez} F.~A.,  {Grand}
  R.~J.,  {Marinacci} F.,   {Pakmor} R.,  2017, \mn@doi [\mnras]
  {10.1093/mnras/stx1301}, \href
  {https://ui.adsabs.harvard.edu/abs/2017MNRAS.470.1259D} {470, 1259}

\bibitem[\protect\citeauthoryear{{Deason}, {Fattahi}, {Belokurov}, {Evans},
  {Grand}, {Marinacci}  \& {Pakmor}}{{Deason} et~al.}{2019}]{Deason:2019}
{Deason} A.~J.,  {Fattahi} A.,  {Belokurov} V.,  {Evans} N.~W.,  {Grand} R.
  J.~J.,  {Marinacci} F.,   {Pakmor} R.,  2019, \mn@doi [\mnras]
  {10.1093/mnras/stz623}, \href
  {https://ui.adsabs.harvard.edu/abs/2019MNRAS.485.3514D} {485, 3514}

\bibitem[\protect\citeauthoryear{{Deason}, {Bose}, {Fattahi}, {Amorisco},
  {Hellwing}  \& {Frenk}}{{Deason} et~al.}{2022}]{Deason:2022}
{Deason} A.~J.,  {Bose} S.,  {Fattahi} A.,  {Amorisco} N.~C.,  {Hellwing} W.,
  {Frenk} C.~S.,  2022, \mn@doi [\mnras] {10.1093/mnras/stab3524}, \href
  {https://ui.adsabs.harvard.edu/abs/2022MNRAS.511.4044D} {511, 4044}

\bibitem[\protect\citeauthoryear{{Diemer} \& {Kravtsov}}{{Diemer} \&
  {Kravtsov}}{2014}]{Deimer:2014}
{Diemer} B.,  {Kravtsov} A.~V.,  2014, \mn@doi [\apj]
  {10.1088/0004-637X/789/1/1}, \href
  {https://ui.adsabs.harvard.edu/abs/2014ApJ...789....1D} {789, 1}

\bibitem[\protect\citeauthoryear{Donlon, Newberg, Sanderson, Bregou, Horta,
  Arora  \& Panithanpaisal}{Donlon et~al.}{2024}]{donlon2024debris}
Donlon T.,  Newberg H.~J.,  Sanderson R.,  Bregou E.,  Horta D.,  Arora A.,
  Panithanpaisal N.,  2024, \mnras, 531, 1422

\bibitem[\protect\citeauthoryear{{Drlica-Wagner} et~al.,}{{Drlica-Wagner}
  et~al.}{2020}]{Drlica-Wagner:2020}
{Drlica-Wagner} A.,  et~al., 2020, \mn@doi [\apj] {10.3847/1538-4357/ab7eb9},
  \href {https://ui.adsabs.harvard.edu/abs/2020ApJ...893...47D} {893, 47}

\bibitem[\protect\citeauthoryear{{Drlica-Wagner} et~al.,}{{Drlica-Wagner}
  et~al.}{2021}]{Drlica-Wagner:2021}
{Drlica-Wagner} A.,  et~al., 2021, \mn@doi [\apjs] {10.3847/1538-4365/ac079d},
  \href {https://ui.adsabs.harvard.edu/abs/2021ApJS..256....2D} {256, 2}

\bibitem[\protect\citeauthoryear{{Dropulic}, {Shipp}, {Kim}, {Mezghanni},
  {Necib}  \& {Lisanti}}{{Dropulic} et~al.}{2024}]{Dropulic:2024}
{Dropulic} A.,  {Shipp} N.,  {Kim} S.,  {Mezghanni} Z.,  {Necib} L.,
  {Lisanti} M.,  2024, \mn@doi [arXiv e-prints] {10.48550/arXiv.2409.13810},
  \href {https://ui.adsabs.harvard.edu/abs/2024arXiv240913810D} {p.
  arXiv:2409.13810}

\bibitem[\protect\citeauthoryear{{Erkal} et~al.,}{{Erkal}
  et~al.}{2018}]{Erkal:2018a}
{Erkal} D.,  et~al., 2018, \mn@doi [\mnras] {10.1093/mnras/sty2518}, \href
  {https://ui.adsabs.harvard.edu/#abs/2018MNRAS.481.3148E} {481, 3148}

\bibitem[\protect\citeauthoryear{{Erkal} et~al.,}{{Erkal}
  et~al.}{2019a}]{Erkal:2019}
{Erkal} D.,  et~al., 2019a, \mn@doi [\mnras] {10.1093/mnras/stz1371}, \href
  {https://ui.adsabs.harvard.edu/abs/2019MNRAS.487.2685E} {487, 2685}

\bibitem[\protect\citeauthoryear{{Erkal} et~al.,}{{Erkal}
  et~al.}{2019b}]{Erkal:2018b}
{Erkal} D.,  et~al., 2019b, \mn@doi [\mnras] {10.1093/mnras/stz1371}, \href
  {https://ui.adsabs.harvard.edu/abs/2019MNRAS.487.2685E} {487, 2685}

\bibitem[\protect\citeauthoryear{{Errani} \& {Navarro}}{{Errani} \&
  {Navarro}}{2021}]{Errani:2021}
{Errani} R.,  {Navarro} J.~F.,  2021, \mn@doi [\mnras]
  {10.1093/mnras/stab1215}, \href
  {https://ui.adsabs.harvard.edu/abs/2021MNRAS.505...18E} {505, 18}

\bibitem[\protect\citeauthoryear{{Errani} \& {Pe{\~n}arrubia}}{{Errani} \&
  {Pe{\~n}arrubia}}{2020}]{Errani:2020}
{Errani} R.,  {Pe{\~n}arrubia} J.,  2020, \mn@doi [\mnras]
  {10.1093/mnras/stz3349}, \href
  {https://ui.adsabs.harvard.edu/abs/2020MNRAS.491.4591E} {491, 4591}

\bibitem[\protect\citeauthoryear{{Errani}, {Pe{\~n}arrubia}, {Laporte}  \&
  {G{\'o}mez}}{{Errani} et~al.}{2017}]{Errani:2017}
{Errani} R.,  {Pe{\~n}arrubia} J.,  {Laporte} C. F.~P.,   {G{\'o}mez} F.~A.,
  2017, \mn@doi [\mnras] {10.1093/mnrasl/slw211}, \href
  {https://ui.adsabs.harvard.edu/abs/2017MNRAS.465L..59E} {465, L59}

\bibitem[\protect\citeauthoryear{{Fattahi}, {Navarro}, {Frenk}, {Oman},
  {Sawala}  \& {Schaller}}{{Fattahi} et~al.}{2018}]{Fattahi:2018}
{Fattahi} A.,  {Navarro} J.~F.,  {Frenk} C.~S.,  {Oman} K.~A.,  {Sawala} T.,
  {Schaller} M.,  2018, \mn@doi [\mnras] {10.1093/mnras/sty408}, \href
  {https://ui.adsabs.harvard.edu/abs/2018MNRAS.476.3816F} {476, 3816}

\bibitem[\protect\citeauthoryear{{Fattahi} et~al.,}{{Fattahi}
  et~al.}{2019}]{Fattahi:2019}
{Fattahi} A.,  et~al., 2019, \mn@doi [\mnras] {10.1093/mnras/stz159}, \href
  {https://ui.adsabs.harvard.edu/abs/2019MNRAS.484.4471F} {484, 4471}

\bibitem[\protect\citeauthoryear{{Fattahi} et~al.,}{{Fattahi}
  et~al.}{2020}]{Fattahi:2020}
{Fattahi} A.,  et~al., 2020, \mn@doi [\mnras] {10.1093/mnras/staa2221}, \href
  {https://ui.adsabs.harvard.edu/abs/2020MNRAS.497.4459F} {497, 4459}

\bibitem[\protect\citeauthoryear{{Faucher-Gigu{\`e}re}, {Lidz}, {Zaldarriaga}
  \& {Hernquist}}{{Faucher-Gigu{\`e}re} et~al.}{2009}]{Faucher-Giguere:2009}
{Faucher-Gigu{\`e}re} C.-A.,  {Lidz} A.,  {Zaldarriaga} M.,   {Hernquist} L.,
  2009, \mn@doi [\apj] {10.1088/0004-637X/703/2/1416}, \href
  {https://ui.adsabs.harvard.edu/abs/2009ApJ...703.1416F} {703, 1416}

\bibitem[\protect\citeauthoryear{{Filion} \& {Wyse}}{{Filion} \&
  {Wyse}}{2021}]{Filion:2021}
{Filion} C.,  {Wyse} R. F.~G.,  2021, \mn@doi [\apj]
  {10.3847/1538-4357/ac2df1}, \href
  {https://ui.adsabs.harvard.edu/abs/2021ApJ...923..218F} {923, 218}

\bibitem[\protect\citeauthoryear{{Garavito-Camargo}, {Besla}, {Laporte},
  {Price-Whelan}, {Cunningham}, {Johnston}, {Weinberg}  \&
  {Gomez}}{{Garavito-Camargo} et~al.}{2020}]{Garavito-Camargo:2020}
{Garavito-Camargo} N.,  {Besla} G.,  {Laporte} C. F.~P.,  {Price-Whelan} A.~M.,
   {Cunningham} E.~C.,  {Johnston} K.~V.,  {Weinberg} M.~D.,   {Gomez} F.~A.,
  2020, arXiv e-prints, \href
  {https://ui.adsabs.harvard.edu/abs/2020arXiv201000816G} {p. arXiv:2010.00816}

\bibitem[\protect\citeauthoryear{{Garrison-Kimmel} et~al.,}{{Garrison-Kimmel}
  et~al.}{2017}]{Garrison-Kimmel:2017}
{Garrison-Kimmel} S.,  et~al., 2017, \mn@doi [\mnras] {10.1093/mnras/stx1710},
  \href {http://adsabs.harvard.edu/abs/2017MNRAS.471.1709G} {471, 1709}

\bibitem[\protect\citeauthoryear{{Garrison-Kimmel} et~al.,}{{Garrison-Kimmel}
  et~al.}{2018}]{Garrison-Kimmel:2018}
{Garrison-Kimmel} S.,  et~al., 2018, preprint, \href
  {https://ui.adsabs.harvard.edu/#abs/2018arXiv180604143G} {p.
  arXiv:1806.04143} (\mn@eprint {arXiv} {1806.04143})

\bibitem[\protect\citeauthoryear{{G{\'o}mez}, {Helmi}, {Cooper}, {Frenk},
  {Navarro}  \& {White}}{{G{\'o}mez} et~al.}{2013}]{Gomez:2013}
{G{\'o}mez} F.~A.,  {Helmi} A.,  {Cooper} A.~P.,  {Frenk} C.~S.,  {Navarro}
  J.~F.,   {White} S. D.~M.,  2013, \mn@doi [\mnras] {10.1093/mnras/stt1838},
  \href {https://ui.adsabs.harvard.edu/abs/2013MNRAS.436.3602G} {436, 3602}

\bibitem[\protect\citeauthoryear{{G{\'o}mez}, {Besla}, {Carpintero},
  {Villalobos}, {O'Shea}  \& {Bell}}{{G{\'o}mez} et~al.}{2015}]{Gomez:2015}
{G{\'o}mez} F.~A.,  {Besla} G.,  {Carpintero} D.~D.,  {Villalobos} {\'A}.,
  {O'Shea} B.~W.,   {Bell} E.~F.,  2015, \mn@doi [\apj]
  {10.1088/0004-637X/802/2/128}, \href
  {https://ui.adsabs.harvard.edu/abs/2015ApJ...802..128G} {802, 128}

\bibitem[\protect\citeauthoryear{{G{\'o}mez}, {White}, {Marinacci}, {Slater},
  {Grand}, {Springel}  \& {Pakmor}}{{G{\'o}mez} et~al.}{2016}]{Gomez:2016}
{G{\'o}mez} F.~A.,  {White} S. D.~M.,  {Marinacci} F.,  {Slater} C.~T.,
  {Grand} R. J.~J.,  {Springel} V.,   {Pakmor} R.,  2016, \mn@doi [\mnras]
  {10.1093/mnras/stv2786}, \href
  {https://ui.adsabs.harvard.edu/abs/2016MNRAS.456.2779G} {456, 2779}

\bibitem[\protect\citeauthoryear{Gommers et~al.,}{Gommers
  et~al.}{2022}]{scipy1.9.1}
Gommers R.,  et~al., 2022, scipy/scipy: SciPy 1.9.1,
  \mn@doi{10.5281/zenodo.7026742}, \url
  {https://doi.org/10.5281/zenodo.7026742}

\bibitem[\protect\citeauthoryear{{Grand} et~al.,}{{Grand}
  et~al.}{2017}]{Grand:2017}
{Grand} R. J.~J.,  et~al., 2017, \mn@doi [\mnras] {10.1093/mnras/stx071}, \href
  {https://ui.adsabs.harvard.edu/abs/2017MNRAS.467..179G} {467, 179}

\bibitem[\protect\citeauthoryear{{Grand} et~al.,}{{Grand}
  et~al.}{2021}]{Grand:2021}
{Grand} R. J.~J.,  et~al., 2021, \mn@doi [\mnras] {10.1093/mnras/stab2492},
  \href {https://ui.adsabs.harvard.edu/abs/2021MNRAS.507.4953G} {507, 4953}

\bibitem[\protect\citeauthoryear{{Grand}, {Fragkoudi}, {G{\'o}mez}, {Jenkins},
  {Marinacci}, {Pakmor}  \& {Springel}}{{Grand} et~al.}{2024}]{Grand:2024}
{Grand} R. J.~J.,  {Fragkoudi} F.,  {G{\'o}mez} F.~A.,  {Jenkins} A.,
  {Marinacci} F.,  {Pakmor} R.,   {Springel} V.,  2024, \mn@doi [\mnras]
  {10.1093/mnras/stae1598}, \href
  {https://ui.adsabs.harvard.edu/abs/2024MNRAS.532.1814G} {532, 1814}

\bibitem[\protect\citeauthoryear{Green, van~den Bosch  \& Jiang}{Green
  et~al.}{2021}]{green2021tidal}
Green S.~B.,  van~den Bosch F.~C.,   Jiang F.,  2021, \mnras, 503, 4075

\bibitem[\protect\citeauthoryear{{Grillmair}}{{Grillmair}}{2006}]{Grillmair:2006}
{Grillmair} C.~J.,  2006, \mn@doi [\apjl] {10.1086/505863}, \href
  {http://adsabs.harvard.edu/abs/2006ApJ...645L..37G} {645, L37}

\bibitem[\protect\citeauthoryear{Grisel et~al.,}{Grisel
  et~al.}{2022}]{scikit-learn1.1.2}
Grisel O.,  et~al., 2022, scikit-learn/scikit-learn: scikit-learn 1.1.2,
  \mn@doi{10.5281/zenodo.6968622}, \url
  {https://doi.org/10.5281/zenodo.6968622}

\bibitem[\protect\citeauthoryear{{Harris} et~al.,}{{Harris}
  et~al.}{2020}]{numpy}
{Harris} C.~R.,  et~al., 2020, \mn@doi [\nat] {10.1038/s41586-020-2649-2},
  \href {https://ui.adsabs.harvard.edu/abs/2020Natur.585..357H} {585, 357}

\bibitem[\protect\citeauthoryear{{He}, {Han}  \& {Li}}{{He}
  et~al.}{2024}]{He:2024}
{He} F.,  {Han} J.,   {Li} Z.,  2024, \mn@doi [arXiv e-prints]
  {10.48550/arXiv.2408.04470}, \href
  {https://ui.adsabs.harvard.edu/abs/2024arXiv240804470H} {p. arXiv:2408.04470}

\bibitem[\protect\citeauthoryear{{Helmi}, {Babusiaux}, {Koppelman}, {Massari},
  {Veljanoski}  \& {Brown}}{{Helmi} et~al.}{2018}]{Helmi:2018}
{Helmi} A.,  {Babusiaux} C.,  {Koppelman} H.~H.,  {Massari} D.,  {Veljanoski}
  J.,   {Brown} A. G.~A.,  2018, \mn@doi [\nat] {10.1038/s41586-018-0625-x},
  \href {https://ui.adsabs.harvard.edu/abs/2018Natur.563...85H} {563, 85}

\bibitem[\protect\citeauthoryear{{Hopkins}}{{Hopkins}}{2015}]{Hopkins:2015}
{Hopkins} P.~F.,  2015, \mn@doi [\mnras] {10.1093/mnras/stv195}, \href
  {https://ui.adsabs.harvard.edu/\#abs/2015MNRAS.450...53H} {450, 53}

\bibitem[\protect\citeauthoryear{{Hopkins} et~al.,}{{Hopkins}
  et~al.}{2018}]{Hopkins:2018}
{Hopkins} P.~F.,  et~al., 2018, \mn@doi [\mnras] {10.1093/mnras/sty1690}, \href
  {https://ui.adsabs.harvard.edu/abs/2018MNRAS.480..800H} {480, 800}

\bibitem[\protect\citeauthoryear{{Hunter}}{{Hunter}}{2007}]{matplotlib}
{Hunter} J.~D.,  2007, \mn@doi [Computing in Science and Engineering]
  {10.1109/MCSE.2007.55}, \href
  {https://ui.adsabs.harvard.edu/abs/2007CSE.....9...90H} {9, 90}

\bibitem[\protect\citeauthoryear{{Ivezi{\'c}} et~al.,}{{Ivezi{\'c}}
  et~al.}{2019}]{LSST:2019}
{Ivezi{\'c}} {\v{Z}}.,  et~al., 2019, \mn@doi [\apj]
  {10.3847/1538-4357/ab042c}, \href
  {https://ui.adsabs.harvard.edu/abs/2019ApJ...873..111I} {873, 111}

\bibitem[\protect\citeauthoryear{{Jenkins}}{{Jenkins}}{2013}]{Jenkins:2013}
{Jenkins} A.,  2013, \mn@doi [\mnras] {10.1093/mnras/stt1154}, \href
  {https://ui.adsabs.harvard.edu/abs/2013MNRAS.434.2094J} {434, 2094}

\bibitem[\protect\citeauthoryear{{Jensen}, {Hayes}, {Sestito}, {McConnachie},
  {Waller}, {Smith}, {Navarro}  \& {Venn}}{{Jensen} et~al.}{2024}]{Jensen:2024}
{Jensen} J.,  {Hayes} C.~R.,  {Sestito} F.,  {McConnachie} A.~W.,  {Waller} F.,
   {Smith} S. E.~T.,  {Navarro} J.,   {Venn} K.~A.,  2024, \mn@doi [\mnras]
  {10.1093/mnras/stad3322}, \href
  {https://ui.adsabs.harvard.edu/abs/2024MNRAS.527.4209J} {527, 4209}

\bibitem[\protect\citeauthoryear{{Ji} et~al.,}{{Ji} et~al.}{2021}]{Ji:2021}
{Ji} A.~P.,  et~al., 2021, arXiv e-prints, \href
  {https://ui.adsabs.harvard.edu/abs/2021arXiv210612656J} {p. arXiv:2106.12656}

\bibitem[\protect\citeauthoryear{{Jung} et~al.,}{{Jung}
  et~al.}{2024}]{AugoraSatellite2024}
{Jung} M.,  et~al., 2024, \mn@doi [\apj] {10.3847/1538-4357/ad245b}, \href
  {https://ui.adsabs.harvard.edu/abs/2024ApJ...964..123J} {964, 123}

\bibitem[\protect\citeauthoryear{{Kallivayalil} et~al.,}{{Kallivayalil}
  et~al.}{2018}]{Kallivayalil:2018}
{Kallivayalil} N.,  et~al., 2018, \mn@doi [\apj] {10.3847/1538-4357/aadfee},
  \href {https://ui.adsabs.harvard.edu/abs/2018ApJ...867...19K} {867, 19}

\bibitem[\protect\citeauthoryear{{Kelley}, {Bullock}, {Garrison-Kimmel},
  {Boylan-Kolchin}, {Pawlowski}  \& {Graus}}{{Kelley}
  et~al.}{2019}]{Kelley:2019}
{Kelley} T.,  {Bullock} J.~S.,  {Garrison-Kimmel} S.,  {Boylan-Kolchin} M.,
  {Pawlowski} M.~S.,   {Graus} A.~S.,  2019, \mn@doi [\mnras]
  {10.1093/mnras/stz1553}, \href
  {https://ui.adsabs.harvard.edu/abs/2019MNRAS.487.4409K} {487, 4409}

\bibitem[\protect\citeauthoryear{{Khoperskov} et~al.,}{{Khoperskov}
  et~al.}{2023}]{Khoperskov:2023}
{Khoperskov} S.,  et~al., 2023, \mn@doi [\aap] {10.1051/0004-6361/202244233},
  \href {https://ui.adsabs.harvard.edu/abs/2023A&A...677A..90K} {677, A90}

\bibitem[\protect\citeauthoryear{{Kim} et~al.,}{{Kim}
  et~al.}{2014}]{Augora2014}
{Kim} J.-h.,  et~al., 2014, \mn@doi [\apjs] {10.1088/0067-0049/210/1/14}, \href
  {https://ui.adsabs.harvard.edu/abs/2014ApJS..210...14K} {210, 14}

\bibitem[\protect\citeauthoryear{Kluyver et~al.,}{Kluyver
  et~al.}{2016}]{jupyter}
Kluyver T.,  et~al., 2016, in Loizides F.,  Scmidt B.,  eds, Positioning and
  Power in Academic Publishing: Players, Agents and Agendas. IOS Press,
  Netherlands, pp 87--90, \url {https://eprints.soton.ac.uk/403913/}

\bibitem[\protect\citeauthoryear{{Koposov}, {Irwin}, {Belokurov},
  {Gonzalez-Solares}, {Yoldas}, {Lewis}, {Metcalfe}  \& {Shanks}}{{Koposov}
  et~al.}{2014}]{Koposov:2014}
{Koposov} S.~E.,  {Irwin} M.,  {Belokurov} V.,  {Gonzalez-Solares} E.,
  {Yoldas} A.~K.,  {Lewis} J.,  {Metcalfe} N.,   {Shanks} T.,  2014, \mn@doi
  [\mnras] {10.1093/mnrasl/slu060}, \href
  {http://adsabs.harvard.edu/abs/2014MNRAS.442L..85K} {442, L85}

\bibitem[\protect\citeauthoryear{{Koposov} et~al.,}{{Koposov}
  et~al.}{2019}]{Koposov:2019}
{Koposov} S.~E.,  et~al., 2019, \mn@doi [\mnras] {10.1093/mnras/stz457}, \href
  {https://ui.adsabs.harvard.edu/abs/2019MNRAS.485.4726K} {485, 4726}

\bibitem[\protect\citeauthoryear{{Koposov} et~al.,}{{Koposov}
  et~al.}{2023}]{Koposov:2023}
{Koposov} S.~E.,  et~al., 2023, \mn@doi [\mnras] {10.1093/mnras/stad551}, \href
  {https://ui.adsabs.harvard.edu/abs/2023MNRAS.521.4936K} {521, 4936}

\bibitem[\protect\citeauthoryear{{Koposov} et~al.,}{{Koposov}
  et~al.}{2024}]{Koposov:2024}
{Koposov} S.~E.,  et~al., 2024, \mn@doi [\mnras] {10.1093/mnras/stae1842},
  \href {https://ui.adsabs.harvard.edu/abs/2024MNRAS.533.1012K} {533, 1012}

\bibitem[\protect\citeauthoryear{{LSST Science Collaboration}}{{LSST Science
  Collaboration}}{2009}]{LSST:2009}
{LSST Science Collaboration} 2009, preprint, \href
  {http://adsabs.harvard.edu/abs/2009arXiv0912.0201L} {} (\mn@eprint {arXiv}
  {0912.0201})

\bibitem[\protect\citeauthoryear{{Li} \& {Helmi}}{{Li} \&
  {Helmi}}{2008}]{Li:2008}
{Li} Y.-S.,  {Helmi} A.,  2008, \mn@doi [\mnras]
  {10.1111/j.1365-2966.2008.12854.x}, \href
  {https://ui.adsabs.harvard.edu/abs/2008MNRAS.385.1365L} {385, 1365}

\bibitem[\protect\citeauthoryear{{Li} et~al.,}{{Li} et~al.}{2021}]{Li:2021}
{Li} T.~S.,  et~al., 2021, \mn@doi [\apj] {10.3847/1538-4357/abeb18}, \href
  {https://ui.adsabs.harvard.edu/abs/2021ApJ...911..149L} {911, 149}

\bibitem[\protect\citeauthoryear{{Li} et~al.,}{{Li} et~al.}{2022}]{Li:2022}
{Li} T.~S.,  et~al., 2022, \mn@doi [\apj] {10.3847/1538-4357/ac46d3}, \href
  {https://ui.adsabs.harvard.edu/abs/2022ApJ...928...30L} {928, 30}

\bibitem[\protect\citeauthoryear{{Lilleengen} et~al.,}{{Lilleengen}
  et~al.}{2023}]{Lilleengen:2023}
{Lilleengen} S.,  et~al., 2023, \mn@doi [\mnras] {10.1093/mnras/stac3108},
  \href {https://ui.adsabs.harvard.edu/abs/2023MNRAS.518..774L} {518, 774}

\bibitem[\protect\citeauthoryear{{Ludlow}, {Schaye}, {Schaller}  \&
  {Richings}}{{Ludlow} et~al.}{2019}]{Ludlow:2019}
{Ludlow} A.~D.,  {Schaye} J.,  {Schaller} M.,   {Richings} J.,  2019, \mn@doi
  [\mnras] {10.1093/mnrasl/slz110}, \href
  {https://ui.adsabs.harvard.edu/abs/2019MNRAS.488L.123L} {488, L123}

\bibitem[\protect\citeauthoryear{{Ludlow}, {Schaye}, {Schaller}  \&
  {Bower}}{{Ludlow} et~al.}{2020}]{Ludlow:2020}
{Ludlow} A.~D.,  {Schaye} J.,  {Schaller} M.,   {Bower} R.,  2020, \mn@doi
  [\mnras] {10.1093/mnras/staa316}, \href
  {https://ui.adsabs.harvard.edu/abs/2020MNRAS.493.2926L} {493, 2926}

\bibitem[\protect\citeauthoryear{{Ludlow}, {Fall}, {Schaye}  \&
  {Obreschkow}}{{Ludlow} et~al.}{2021}]{Ludlow:2021}
{Ludlow} A.~D.,  {Fall} S.~M.,  {Schaye} J.,   {Obreschkow} D.,  2021, \mn@doi
  [\mnras] {10.1093/mnras/stab2770}, \href
  {https://ui.adsabs.harvard.edu/abs/2021MNRAS.508.5114L} {508, 5114}

\bibitem[\protect\citeauthoryear{{Ludlow}, {Fall}, {Wilkinson}, {Schaye}  \&
  {Obreschkow}}{{Ludlow} et~al.}{2023}]{Ludlow:2023}
{Ludlow} A.~D.,  {Fall} S.~M.,  {Wilkinson} M.~J.,  {Schaye} J.,   {Obreschkow}
  D.,  2023, \mn@doi [\mnras] {10.1093/mnras/stad2615}, \href
  {https://ui.adsabs.harvard.edu/abs/2023MNRAS.525.5614L} {525, 5614}

\bibitem[\protect\citeauthoryear{{Malhan} \& {Ibata}}{{Malhan} \&
  {Ibata}}{2018}]{Malhan:2018a}
{Malhan} K.,  {Ibata} R.~A.,  2018, \mn@doi [\mnras] {10.1093/mnras/sty912},
  \href {https://ui.adsabs.harvard.edu/abs/2018MNRAS.477.4063M} {477, 4063}

\bibitem[\protect\citeauthoryear{{Marinacci}, {Pakmor}  \&
  {Springel}}{{Marinacci} et~al.}{2014}]{Marinacci:2014}
{Marinacci} F.,  {Pakmor} R.,   {Springel} V.,  2014, \mn@doi [\mnras]
  {10.1093/mnras/stt2003}, \href
  {https://ui.adsabs.harvard.edu/abs/2014MNRAS.437.1750M} {437, 1750}

\bibitem[\protect\citeauthoryear{{Marinacci}, {Grand}, {Pakmor}, {Springel},
  {G{\'o}mez}, {Frenk}  \& {White}}{{Marinacci} et~al.}{2017}]{Marinacci:2017}
{Marinacci} F.,  {Grand} R. J.~J.,  {Pakmor} R.,  {Springel} V.,  {G{\'o}mez}
  F.~A.,  {Frenk} C.~S.,   {White} S. D.~M.,  2017, \mn@doi [\mnras]
  {10.1093/mnras/stw3366}, \href
  {https://ui.adsabs.harvard.edu/abs/2017MNRAS.466.3859M} {466, 3859}

\bibitem[\protect\citeauthoryear{{Mateu}}{{Mateu}}{2023}]{Mateu:2023}
{Mateu} C.,  2023, \mn@doi [\mnras] {10.1093/mnras/stad321}, \href
  {https://ui.adsabs.harvard.edu/abs/2023MNRAS.520.5225M} {520, 5225}

\bibitem[\protect\citeauthoryear{{M}c{K}inney}{{M}c{K}inney}{2010}]{pandas}
{M}c{K}inney W.,  2010, in {S}t\'efan van~der {W}alt {J}arrod {M}illman eds,
  {P}roceedings of the 9th {P}ython in {S}cience {C}onference. pp 56 -- 61,
  \mn@doi{10.25080/Majora-92bf1922-00a}

\bibitem[\protect\citeauthoryear{{Monachesi}, {G{\'o}mez}, {Grand},
  {Kauffmann}, {Marinacci}, {Pakmor}, {Springel}  \& {Frenk}}{{Monachesi}
  et~al.}{2016}]{Monachesi:2016}
{Monachesi} A.,  {G{\'o}mez} F.~A.,  {Grand} R. J.~J.,  {Kauffmann} G.,
  {Marinacci} F.,  {Pakmor} R.,  {Springel} V.,   {Frenk} C.~S.,  2016, \mn@doi
  [\mnras] {10.1093/mnrasl/slw052}, \href
  {https://ui.adsabs.harvard.edu/abs/2016MNRAS.459L..46M} {459, L46}

\bibitem[\protect\citeauthoryear{{Monachesi} et~al.,}{{Monachesi}
  et~al.}{2019}]{Monachesi:2019}
{Monachesi} A.,  et~al., 2019, \mn@doi [\mnras] {10.1093/mnras/stz538}, \href
  {https://ui.adsabs.harvard.edu/abs/2019MNRAS.485.2589M} {485, 2589}

\bibitem[\protect\citeauthoryear{{Mori}, {Di Matteo}, {Salvadori},
  {Khoperskov}, {Pagnini}  \& {Haywood}}{{Mori} et~al.}{2024}]{Mori:2024}
{Mori} A.,  {Di Matteo} P.,  {Salvadori} S.,  {Khoperskov} S.,  {Pagnini} G.,
  {Haywood} M.,  2024, \mn@doi [arXiv e-prints] {10.48550/arXiv.2401.13737},
  \href {https://ui.adsabs.harvard.edu/abs/2024arXiv240113737M} {p.
  arXiv:2401.13737}

\bibitem[\protect\citeauthoryear{{Munshi}, {Brooks}, {Applebaum},
  {Christensen}, {Quinn}  \& {Sligh}}{{Munshi} et~al.}{2021}]{Munshi:2021}
{Munshi} F.,  {Brooks} A.~M.,  {Applebaum} E.,  {Christensen} C.~R.,  {Quinn}
  T.,   {Sligh} S.,  2021, \mn@doi [\apj] {10.3847/1538-4357/ac0db6}, \href
  {https://ui.adsabs.harvard.edu/abs/2021ApJ...923...35M} {923, 35}

\bibitem[\protect\citeauthoryear{{Nadler}, {Mao}, {Wechsler}, {Garrison-Kimmel}
   \& {Wetzel}}{{Nadler} et~al.}{2018}]{Nadler:2018}
{Nadler} E.~O.,  {Mao} Y.-Y.,  {Wechsler} R.~H.,  {Garrison-Kimmel} S.,
  {Wetzel} A.,  2018, \mn@doi [\apj] {10.3847/1538-4357/aac266}, \href
  {https://ui.adsabs.harvard.edu/abs/2018ApJ...859..129N} {859, 129}

\bibitem[\protect\citeauthoryear{{Nadler} et~al.,}{{Nadler}
  et~al.}{2020}]{Nadler:2020}
{Nadler} E.~O.,  et~al., 2020, \mn@doi [\apj] {10.3847/1538-4357/ab846a}, \href
  {https://ui.adsabs.harvard.edu/abs/2020ApJ...893...48N} {893, 48}

\bibitem[\protect\citeauthoryear{{Nadler}, {Birrer}, {Gilman}, {Wechsler},
  {Du}, {Benson}, {Nierenberg}  \& {Treu}}{{Nadler} et~al.}{2021}]{Nadler:2021}
{Nadler} E.~O.,  {Birrer} S.,  {Gilman} D.,  {Wechsler} R.~H.,  {Du} X.,
  {Benson} A.,  {Nierenberg} A.~M.,   {Treu} T.,  2021, \mn@doi [\apj]
  {10.3847/1538-4357/abf9a3}, \href
  {https://ui.adsabs.harvard.edu/abs/2021ApJ...917....7N} {917, 7}

\bibitem[\protect\citeauthoryear{{Naidu}, {Conroy}, {Bonaca}, {Johnson},
  {Ting}, {Caldwell}, {Zaritsky}  \& {Cargile}}{{Naidu}
  et~al.}{2020}]{Naidu:2020}
{Naidu} R.~P.,  {Conroy} C.,  {Bonaca} A.,  {Johnson} B.~D.,  {Ting} Y.-S.,
  {Caldwell} N.,  {Zaritsky} D.,   {Cargile} P.~A.,  2020, \mn@doi [\apj]
  {10.3847/1538-4357/abaef4}, \href
  {https://ui.adsabs.harvard.edu/abs/2020ApJ...901...48N} {901, 48}

\bibitem[\protect\citeauthoryear{{Oman} \& {Riley}}{{Oman} \&
  {Riley}}{2024}]{Oman:2024}
{Oman} K.~A.,  {Riley} A.~H.,  2024, \mn@doi [\mnras] {10.1093/mnrasl/slae042},
  \href {https://ui.adsabs.harvard.edu/abs/2024MNRAS.532L..48O} {532, L48}

\bibitem[\protect\citeauthoryear{{Ou}, {Eilers}, {Necib}  \& {Frebel}}{{Ou}
  et~al.}{2024a}]{Ou:2024rotcurve}
{Ou} X.,  {Eilers} A.-C.,  {Necib} L.,   {Frebel} A.,  2024a, \mn@doi [\mnras]
  {10.1093/mnras/stae034}, \href
  {https://ui.adsabs.harvard.edu/abs/2024MNRAS.528..693O} {528, 693}

\bibitem[\protect\citeauthoryear{{Ou} et~al.,}{{Ou}
  et~al.}{2024b}]{Ou:2024hercules}
{Ou} X.,  et~al., 2024b, \mn@doi [\apj] {10.3847/1538-4357/ad2f27}, \href
  {https://ui.adsabs.harvard.edu/abs/2024ApJ...966...33O} {966, 33}

\bibitem[\protect\citeauthoryear{{Pace}, {Erkal}  \& {Li}}{{Pace}
  et~al.}{2022}]{Pace:2022}
{Pace} A.~B.,  {Erkal} D.,   {Li} T.~S.,  2022, arXiv e-prints, \href
  {https://ui.adsabs.harvard.edu/abs/2022arXiv220505699P} {p. arXiv:2205.05699}

\bibitem[\protect\citeauthoryear{{Pakmor}, {Springel}, {Bauer}, {Mocz},
  {Munoz}, {Ohlmann}, {Schaal}  \& {Zhu}}{{Pakmor} et~al.}{2016}]{Pakmor:2016}
{Pakmor} R.,  {Springel} V.,  {Bauer} A.,  {Mocz} P.,  {Munoz} D.~J.,
  {Ohlmann} S.~T.,  {Schaal} K.,   {Zhu} C.,  2016, \mn@doi [\mnras]
  {10.1093/mnras/stv2380}, \href
  {https://ui.adsabs.harvard.edu/abs/2016MNRAS.455.1134P} {455, 1134}

\bibitem[\protect\citeauthoryear{{Pakmor} et~al.,}{{Pakmor}
  et~al.}{2017}]{Pakmor:2017}
{Pakmor} R.,  et~al., 2017, \mn@doi [\mnras] {10.1093/mnras/stx1074}, \href
  {https://ui.adsabs.harvard.edu/abs/2017MNRAS.469.3185P} {469, 3185}

\bibitem[\protect\citeauthoryear{{Panithanpaisal}, {Sanderson}, {Wetzel},
  {Cunningham}, {Bailin}  \& {Faucher-Gigu{\`e}re}}{{Panithanpaisal}
  et~al.}{2021}]{Panithanpaisal:2021}
{Panithanpaisal} N.,  {Sanderson} R.~E.,  {Wetzel} A.,  {Cunningham} E.~C.,
  {Bailin} J.,   {Faucher-Gigu{\`e}re} C.-A.,  2021, \mn@doi [\apj]
  {10.3847/1538-4357/ac1109}, \href
  {https://ui.adsabs.harvard.edu/abs/2021ApJ...920...10P} {920, 10}

\bibitem[\protect\citeauthoryear{{Patel} et~al.,}{{Patel}
  et~al.}{2020}]{Patel:2020}
{Patel} E.,  et~al., 2020, \mn@doi [\apj] {10.3847/1538-4357/ab7b75}, \href
  {https://ui.adsabs.harvard.edu/abs/2020ApJ...893..121P} {893, 121}

\bibitem[\protect\citeauthoryear{Pedregosa et~al.,}{Pedregosa
  et~al.}{2011}]{scikit-learn}
Pedregosa F.,  et~al., 2011, Journal of Machine Learning Research, 12, 2825

\bibitem[\protect\citeauthoryear{{Perez} \& {Granger}}{{Perez} \&
  {Granger}}{2007}]{ipython}
{Perez} F.,  {Granger} B.~E.,  2007, \mn@doi [Computing in Science and
  Engineering] {10.1109/MCSE.2007.53}, \href
  {https://ui.adsabs.harvard.edu/abs/2007CSE.....9c..21P} {9, 21}

\bibitem[\protect\citeauthoryear{{Planck Collaboration} et~al.,}{{Planck
  Collaboration} et~al.}{2014}]{PlanckCollaboration:2014}
{Planck Collaboration} et~al., 2014, \mn@doi [\aap]
  {10.1051/0004-6361/201321591}, \href
  {https://ui.adsabs.harvard.edu/abs/2014A&A...571A..16P} {571, A16}

\bibitem[\protect\citeauthoryear{{Pozo}, {Broadhurst}, {de Martino}, {Chiueh},
  {Smoot}, {Bonoli}  \& {Angulo}}{{Pozo} et~al.}{2024}]{Pozo:2024}
{Pozo} A.,  {Broadhurst} T.,  {de Martino} I.,  {Chiueh} T.,  {Smoot} G.~F.,
  {Bonoli} S.,   {Angulo} R.,  2024, \mn@doi [\prd]
  {10.1103/PhysRevD.110.043534}, \href
  {https://ui.adsabs.harvard.edu/abs/2024PhRvD.110d3534P} {110, 043534}

\bibitem[\protect\citeauthoryear{{Qi}, {Zivick}, {Pace}, {Riley}  \&
  {Strigari}}{{Qi} et~al.}{2022}]{Qi:2022}
{Qi} Y.,  {Zivick} P.,  {Pace} A.~B.,  {Riley} A.~H.,   {Strigari} L.~E.,
  2022, \mn@doi [\mnras] {10.1093/mnras/stac805}, \href
  {https://ui.adsabs.harvard.edu/abs/2022MNRAS.512.5601Q} {512, 5601}

\bibitem[\protect\citeauthoryear{{Racca} et~al.,}{{Racca}
  et~al.}{2016}]{Racca:2016}
{Racca} G.~D.,  et~al., 2016, in {MacEwen} H.~A.,  {Fazio} G.~G.,  {Lystrup}
  M.,  {Batalha} N.,  {Siegler} N.,   {Tong} E.~C.,  eds,  Society of
  Photo-Optical Instrumentation Engineers (SPIE) Conference Series Vol. 9904,
  Space Telescopes and Instrumentation 2016: Optical, Infrared, and Millimeter
  Wave. p. 99040O (\mn@eprint {arXiv} {1610.05508}),
  \mn@doi{10.1117/12.2230762}

\bibitem[\protect\citeauthoryear{{Revaz} \& {Jablonka}}{{Revaz} \&
  {Jablonka}}{2018}]{Revaz:2018}
{Revaz} Y.,  {Jablonka} P.,  2018, \mn@doi [\aap]
  {10.1051/0004-6361/201832669}, \href
  {https://ui.adsabs.harvard.edu/abs/2018A&A...616A..96R} {616, A96}

\bibitem[\protect\citeauthoryear{{Richings} et~al.,}{{Richings}
  et~al.}{2020}]{Richings:2020}
{Richings} J.,  et~al., 2020, \mn@doi [\mnras] {10.1093/mnras/stz3448}, \href
  {https://ui.adsabs.harvard.edu/abs/2020MNRAS.492.5780R} {492, 5780}

\bibitem[\protect\citeauthoryear{{Riley} et~al.,}{{Riley}
  et~al.}{2019}]{Riley:2019}
{Riley} A.~H.,  et~al., 2019, \mn@doi [\mnras] {10.1093/mnras/stz973}, \href
  {https://ui.adsabs.harvard.edu/abs/2019MNRAS.486.2679R} {486, 2679}

\bibitem[\protect\citeauthoryear{{Rocha}, {Peter}  \& {Bullock}}{{Rocha}
  et~al.}{2012}]{Rocha:2012}
{Rocha} M.,  {Peter} A. H.~G.,   {Bullock} J.,  2012, \mn@doi [\mnras]
  {10.1111/j.1365-2966.2012.21432.x}, \href
  {https://ui.adsabs.harvard.edu/abs/2012MNRAS.425..231R} {425, 231}

\bibitem[\protect\citeauthoryear{{Sales}, {Wetzel}  \& {Fattahi}}{{Sales}
  et~al.}{2022}]{Sales:2022}
{Sales} L.~V.,  {Wetzel} A.,   {Fattahi} A.,  2022, \mn@doi [Nature Astronomy]
  {10.1038/s41550-022-01689-w}, \href
  {https://ui.adsabs.harvard.edu/abs/2022NatAs.tmp..130S} {}

\bibitem[\protect\citeauthoryear{{Samuel} et~al.,}{{Samuel}
  et~al.}{2020}]{Samuel:2020}
{Samuel} J.,  et~al., 2020, \mn@doi [\mnras] {10.1093/mnras/stz3054}, \href
  {https://ui.adsabs.harvard.edu/abs/2020MNRAS.491.1471S} {491, 1471}

\bibitem[\protect\citeauthoryear{Sanders, Lilley, Vasiliev, Evans  \&
  Erkal}{Sanders et~al.}{2020}]{Sanders:2020}
Sanders J.~L.,  Lilley E.~J.,  Vasiliev E.,  Evans N.~W.,   Erkal D.,  2020,
  Monthly Notices of the Royal Astronomical Society, 499, 4793

\bibitem[\protect\citeauthoryear{{Santistevan}, {Wetzel}, {Tollerud},
  {Sanderson}  \& {Samuel}}{{Santistevan} et~al.}{2023}]{Santistevan:2023}
{Santistevan} I.~B.,  {Wetzel} A.,  {Tollerud} E.,  {Sanderson} R.~E.,
  {Samuel} J.,  2023, \mn@doi [\mnras] {10.1093/mnras/stac3100}, \href
  {https://ui.adsabs.harvard.edu/abs/2023MNRAS.518.1427S} {518, 1427}

\bibitem[\protect\citeauthoryear{{Santistevan}, {Wetzel}, {Tollerud},
  {Sanderson}, {Moreno}  \& {Patel}}{{Santistevan}
  et~al.}{2024}]{Santistevan:2024}
{Santistevan} I.~B.,  {Wetzel} A.,  {Tollerud} E.,  {Sanderson} R.~E.,
  {Moreno} J.,   {Patel} E.,  2024, \mn@doi [\mnras] {10.1093/mnras/stad3757},
  \href {https://ui.adsabs.harvard.edu/abs/2024MNRAS.527.8841S} {527, 8841}

\bibitem[\protect\citeauthoryear{{Santos-Santos}, {Fattahi}, {Sales}  \&
  {Navarro}}{{Santos-Santos} et~al.}{2021}]{Santos-Santos:2021}
{Santos-Santos} I. M.~E.,  {Fattahi} A.,  {Sales} L.~V.,   {Navarro} J.~F.,
  2021, \mn@doi [\mnras] {10.1093/mnras/stab1020}, \href
  {https://ui.adsabs.harvard.edu/abs/2021MNRAS.504.4551S} {504, 4551}

\bibitem[\protect\citeauthoryear{{Sawala} et~al.,}{{Sawala}
  et~al.}{2016}]{Sawala:2016}
{Sawala} T.,  et~al., 2016, \mn@doi [\mnras] {10.1093/mnras/stw145}, \href
  {https://ui.adsabs.harvard.edu/abs/2016MNRAS.457.1931S} {457, 1931}

\bibitem[\protect\citeauthoryear{{Schaye} et~al.,}{{Schaye}
  et~al.}{2015}]{Schaye:2015}
{Schaye} J.,  et~al., 2015, \mn@doi [\mnras] {10.1093/mnras/stu2058}, \href
  {https://ui.adsabs.harvard.edu/abs/2015MNRAS.446..521S} {446, 521}

\bibitem[\protect\citeauthoryear{{Searle} \& {Zinn}}{{Searle} \&
  {Zinn}}{1978}]{Searle:1978}
{Searle} L.,  {Zinn} R.,  1978, \mn@doi [\apj] {10.1086/156499}, \href
  {https://ui.adsabs.harvard.edu/#abs/1978ApJ...225..357S} {225, 357}

\bibitem[\protect\citeauthoryear{{Sestito} et~al.,}{{Sestito}
  et~al.}{2023}]{Sestito:2023}
{Sestito} F.,  et~al., 2023, \mn@doi [\mnras] {10.1093/mnras/stad2427}, \href
  {https://ui.adsabs.harvard.edu/abs/2023MNRAS.525.2875S} {525, 2875}

\bibitem[\protect\citeauthoryear{{Shipp} et~al.,}{{Shipp}
  et~al.}{2018}]{Shipp:2018}
{Shipp} N.,  et~al., 2018, \mn@doi [\apj] {10.3847/1538-4357/aacdab}, \href
  {https://ui.adsabs.harvard.edu/\#abs/2018ApJ...862..114S} {862, 114}

\bibitem[\protect\citeauthoryear{{Shipp} et~al.,}{{Shipp}
  et~al.}{2021}]{Shipp:2021}
{Shipp} N.,  et~al., 2021, \mn@doi [\apj] {10.3847/1538-4357/ac2e93}, \href
  {https://ui.adsabs.harvard.edu/abs/2021ApJ...923..149S} {923, 149}

\bibitem[\protect\citeauthoryear{{Shipp} et~al.,}{{Shipp}
  et~al.}{2023}]{Shipp:2023}
{Shipp} N.,  et~al., 2023, \mn@doi [\apj] {10.3847/1538-4357/acc582}, \href
  {https://ui.adsabs.harvard.edu/abs/2023ApJ...949...44S} {949, 44}

\bibitem[\protect\citeauthoryear{{Simpson}, {Grand}, {G{\'o}mez}, {Marinacci},
  {Pakmor}, {Springel}, {Campbell}  \& {Frenk}}{{Simpson}
  et~al.}{2018}]{Simpson:2018}
{Simpson} C.~M.,  {Grand} R. J.~J.,  {G{\'o}mez} F.~A.,  {Marinacci} F.,
  {Pakmor} R.,  {Springel} V.,  {Campbell} D. J.~R.,   {Frenk} C.~S.,  2018,
  \mn@doi [\mnras] {10.1093/mnras/sty774}, \href
  {https://ui.adsabs.harvard.edu/abs/2018MNRAS.478..548S} {478, 548}

\bibitem[\protect\citeauthoryear{{Spergel} et~al.,}{{Spergel}
  et~al.}{2013}]{Spergel:2013}
{Spergel} D.,  et~al., 2013, arXiv e-prints, \href
  {https://ui.adsabs.harvard.edu/abs/2013arXiv1305.5422S} {p. arXiv:1305.5422}

\bibitem[\protect\citeauthoryear{{Springel}}{{Springel}}{2005}]{Springel:2005}
{Springel} V.,  2005, \mn@doi [\mnras] {10.1111/j.1365-2966.2005.09655.x},
  \href {https://ui.adsabs.harvard.edu/abs/2005MNRAS.364.1105S} {364, 1105}

\bibitem[\protect\citeauthoryear{{Springel}}{{Springel}}{2010}]{Springel:2010}
{Springel} V.,  2010, \mn@doi [\mnras] {10.1111/j.1365-2966.2009.15715.x},
  \href {https://ui.adsabs.harvard.edu/abs/2010MNRAS.401..791S} {401, 791}

\bibitem[\protect\citeauthoryear{{Springel} \& {Hernquist}}{{Springel} \&
  {Hernquist}}{2003}]{Springel:2003}
{Springel} V.,  {Hernquist} L.,  2003, \mn@doi [\mnras]
  {10.1046/j.1365-8711.2003.06206.x}, \href
  {https://ui.adsabs.harvard.edu/abs/2003MNRAS.339..289S} {339, 289}

\bibitem[\protect\citeauthoryear{{Springel}, {White}, {Tormen}  \&
  {Kauffmann}}{{Springel} et~al.}{2001}]{Springel:2001}
{Springel} V.,  {White} S. D.~M.,  {Tormen} G.,   {Kauffmann} G.,  2001,
  \mn@doi [\mnras] {10.1046/j.1365-8711.2001.04912.x}, \href
  {https://ui.adsabs.harvard.edu/abs/2001MNRAS.328..726S} {328, 726}

\bibitem[\protect\citeauthoryear{{Springel}, {Di Matteo}  \&
  {Hernquist}}{{Springel} et~al.}{2005}]{Springel:2005feedback}
{Springel} V.,  {Di Matteo} T.,   {Hernquist} L.,  2005, \mn@doi [\mnras]
  {10.1111/j.1365-2966.2005.09238.x}, \href
  {https://ui.adsabs.harvard.edu/abs/2005MNRAS.361..776S} {361, 776}

\bibitem[\protect\citeauthoryear{{Starkenburg} et~al.,}{{Starkenburg}
  et~al.}{2017}]{Starkenburg:2017}
{Starkenburg} E.,  et~al., 2017, \mn@doi [\mnras] {10.1093/mnras/stx1068},
  \href {https://ui.adsabs.harvard.edu/abs/2017MNRAS.471.2587S} {471, 2587}

\bibitem[\protect\citeauthoryear{{Tarumi}, {Yoshida}  \& {Frebel}}{{Tarumi}
  et~al.}{2021}]{Tarumi:2021}
{Tarumi} Y.,  {Yoshida} N.,   {Frebel} A.,  2021, \mn@doi [\apjl]
  {10.3847/2041-8213/ac024e}, \href
  {https://ui.adsabs.harvard.edu/abs/2021ApJ...914L..10T} {914, L10}

\bibitem[\protect\citeauthoryear{{Torrealba} et~al.,}{{Torrealba}
  et~al.}{2019}]{Torrealba:2019}
{Torrealba} G.,  et~al., 2019, \mn@doi [\mnras] {10.1093/mnras/stz1624}, \href
  {https://ui.adsabs.harvard.edu/abs/2019MNRAS.488.2743T} {488, 2743}

\bibitem[\protect\citeauthoryear{{Vasiliev}}{{Vasiliev}}{2019a}]{Vasiliev:2019}
{Vasiliev} E.,  2019a, \mn@doi [\mnras] {10.1093/mnras/sty2672}, \href
  {https://ui.adsabs.harvard.edu/abs/2019MNRAS.482.1525V} {482, 1525}

\bibitem[\protect\citeauthoryear{{Vasiliev}}{{Vasiliev}}{2019b}]{agama}
{Vasiliev} E.,  2019b, \mn@doi [\mnras] {10.1093/mnras/sty2672}, \href
  {https://ui.adsabs.harvard.edu/abs/2019MNRAS.482.1525V} {482, 1525}

\bibitem[\protect\citeauthoryear{{Vasiliev}}{{Vasiliev}}{2024}]{Vasiliev:2024}
{Vasiliev} E.,  2024, \mn@doi [\mnras] {10.1093/mnras/stad2612}, \href
  {https://ui.adsabs.harvard.edu/abs/2024MNRAS.527..437V} {527, 437}

\bibitem[\protect\citeauthoryear{{Vasiliev}, {Belokurov}  \&
  {Erkal}}{{Vasiliev} et~al.}{2021}]{Vasiliev:2021}
{Vasiliev} E.,  {Belokurov} V.,   {Erkal} D.,  2021, \mn@doi [\mnras]
  {10.1093/mnras/staa3673}, \href
  {https://ui.adsabs.harvard.edu/abs/2021MNRAS.501.2279V} {501, 2279}

\bibitem[\protect\citeauthoryear{{Vera-Casanova} et~al.,}{{Vera-Casanova}
  et~al.}{2022}]{Vera-Casanova:2022}
{Vera-Casanova} A.,  et~al., 2022, \mn@doi [\mnras] {10.1093/mnras/stac1636},
  \href {https://ui.adsabs.harvard.edu/abs/2022MNRAS.514.4898V} {514, 4898}

\bibitem[\protect\citeauthoryear{{Vienneau}, {Evans}, {Hartl}, {Bozorgnia},
  {Strigari}, {Riley}  \& {Shipp}}{{Vienneau} et~al.}{2024}]{Vienneau:2024}
{Vienneau} E.,  {Evans} A.~J.,  {Hartl} O.~V.,  {Bozorgnia} N.,  {Strigari}
  L.~E.,  {Riley} A.~H.,   {Shipp} N.,  2024, \mn@doi [arXiv e-prints]
  {10.48550/arXiv.2403.15544}, \href
  {https://ui.adsabs.harvard.edu/abs/2024arXiv240315544V} {p. arXiv:2403.15544}

\bibitem[\protect\citeauthoryear{{Virtanen} et~al.,}{{Virtanen}
  et~al.}{2020}]{scipy}
{Virtanen} P.,  et~al., 2020, \mn@doi [Nature Methods]
  {10.1038/s41592-019-0686-2}, \href
  {https://ui.adsabs.harvard.edu/abs/2020NatMe..17..261V} {17, 261}

\bibitem[\protect\citeauthoryear{{Vivas}, {Mart{\'\i}nez-V{\'a}zquez},
  {Walker}, {Belokurov}, {Li}  \& {Erkal}}{{Vivas} et~al.}{2022}]{Vivas:2022}
{Vivas} A.~K.,  {Mart{\'\i}nez-V{\'a}zquez} C.~E.,  {Walker} A.~R.,
  {Belokurov} V.,  {Li} T.~S.,   {Erkal} D.,  2022, \mn@doi [\apj]
  {10.3847/1538-4357/ac43bd}, \href
  {https://ui.adsabs.harvard.edu/abs/2022ApJ...926...78V} {926, 78}

\bibitem[\protect\citeauthoryear{{Vogelsberger}, {Genel}, {Sijacki}, {Torrey},
  {Springel}  \& {Hernquist}}{{Vogelsberger} et~al.}{2013}]{Vogelsberger:2013}
{Vogelsberger} M.,  {Genel} S.,  {Sijacki} D.,  {Torrey} P.,  {Springel} V.,
  {Hernquist} L.,  2013, \mn@doi [\mnras] {10.1093/mnras/stt1789}, \href
  {https://ui.adsabs.harvard.edu/abs/2013MNRAS.436.3031V} {436, 3031}

\bibitem[\protect\citeauthoryear{{Wagg} \& {Broekgaarden}}{{Wagg} \&
  {Broekgaarden}}{2024}]{software-citation-station-paper}
{Wagg} T.,  {Broekgaarden} F.~S.,  2024, arXiv e-prints, \href
  {https://ui.adsabs.harvard.edu/abs/2024arXiv240604405W} {p. arXiv:2406.04405}

\bibitem[\protect\citeauthoryear{Wagg, Broekgaarden  \& Gültekin}{Wagg
  et~al.}{2024}]{software-citation-station-zenodo}
Wagg T.,  Broekgaarden F.,   Gültekin K.,  2024,
  TomWagg/software-citation-station: v1.2, \mn@doi{10.5281/zenodo.13225824},
  \url {https://doi.org/10.5281/zenodo.13225824}

\bibitem[\protect\citeauthoryear{{Wang}, {Han}, {Cautun}, {Li}  \&
  {Ishigaki}}{{Wang} et~al.}{2020}]{Wang:2020}
{Wang} W.,  {Han} J.,  {Cautun} M.,  {Li} Z.,   {Ishigaki} M.~N.,  2020,
  \mn@doi [Science China Physics, Mechanics, and Astronomy]
  {10.1007/s11433-019-1541-6}, \href
  {https://ui.adsabs.harvard.edu/abs/2020SCPMA..6309801W} {63, 109801}

\bibitem[\protect\citeauthoryear{{Wang} et~al.,}{{Wang}
  et~al.}{2022}]{Wang:2022}
{Wang} W.,  et~al., 2022, \mn@doi [\apj] {10.3847/1538-4357/ac9b19}, \href
  {https://ui.adsabs.harvard.edu/abs/2022ApJ...941..108W} {941, 108}

\bibitem[\protect\citeauthoryear{{Wang}, {Mansfield}, {Nadler}, {Darragh-Ford},
  {Wechsler}, {Yang}  \& {Yu}}{{Wang} et~al.}{2024}]{Wang:2024}
{Wang} Y.,  {Mansfield} P.,  {Nadler} E.~O.,  {Darragh-Ford} E.,  {Wechsler}
  R.~H.,  {Yang} D.,   {Yu} H.-B.,  2024, \mn@doi [arXiv e-prints]
  {10.48550/arXiv.2408.01487}, \href
  {https://ui.adsabs.harvard.edu/abs/2024arXiv240801487W} {p. arXiv:2408.01487}

\bibitem[\protect\citeauthoryear{{Wetzel}, {Deason}  \&
  {Garrison-Kimmel}}{{Wetzel} et~al.}{2015}]{Wetzel:2015}
{Wetzel} A.~R.,  {Deason} A.~J.,   {Garrison-Kimmel} S.,  2015, \mn@doi [\apj]
  {10.1088/0004-637X/807/1/49}, \href
  {https://ui.adsabs.harvard.edu/abs/2015ApJ...807...49W} {807, 49}

\bibitem[\protect\citeauthoryear{{Wetzel}, {Hopkins}, {Kim},
  {Faucher-Gigu{\`e}re}, {Kere{\v{s}}}  \& {Quataert}}{{Wetzel}
  et~al.}{2016}]{Wetzel:2016}
{Wetzel} A.~R.,  {Hopkins} P.~F.,  {Kim} J.-h.,  {Faucher-Gigu{\`e}re} C.-A.,
  {Kere{\v{s}}} D.,   {Quataert} E.,  2016, \mn@doi [\apj]
  {10.3847/2041-8205/827/2/L23}, \href
  {https://ui.adsabs.harvard.edu/#abs/2016ApJ...827L..23W} {827, L23}

\bibitem[\protect\citeauthoryear{{Wetzel} et~al.,}{{Wetzel}
  et~al.}{2023}]{Wetzel:2023}
{Wetzel} A.,  et~al., 2023, \mn@doi [\apjs] {10.3847/1538-4365/acb99a}, \href
  {https://ui.adsabs.harvard.edu/abs/2023ApJS..265...44W} {265, 44}

\bibitem[\protect\citeauthoryear{{York} et~al.,}{{York}
  et~al.}{2000}]{York:2000}
{York} D.~G.,  et~al., 2000, \mn@doi [\aj] {10.1086/301513}, \href
  {http://adsabs.harvard.edu/abs/2000AJ....120.1579Y} {120, 1579}

\bibitem[\protect\citeauthoryear{{de Jong} et~al.,}{{de Jong}
  et~al.}{2019}]{4MOST:2019}
{de Jong} R.~S.,  et~al., 2019, \mn@doi [The Messenger]
  {10.18727/0722-6691/5117}, \href
  {https://ui.adsabs.harvard.edu/abs/2019Msngr.175....3D} {175, 3}

\bibitem[\protect\citeauthoryear{pandas~development team}{pandas~development
  team}{2022}]{pandas1.5.0}
pandas~development team T.,  2022, pandas-dev/pandas: Pandas,
  \mn@doi{10.5281/zenodo.7093122}, \url
  {https://doi.org/10.5281/zenodo.7093122}

\bibitem[\protect\citeauthoryear{{van den Bosch} \& {Ogiya}}{{van den Bosch} \&
  {Ogiya}}{2018}]{vandenBosch:2018}
{van den Bosch} F.~C.,  {Ogiya} G.,  2018, \mn@doi [\mnras]
  {10.1093/mnras/sty084}, \href
  {https://ui.adsabs.harvard.edu/abs/2018MNRAS.475.4066V} {475, 4066}

\bibitem[\protect\citeauthoryear{{van der Velden}}{{van der
  Velden}}{2020}]{cmasher}
{van der Velden} E.,  2020, \mn@doi [The Journal of Open Source Software]
  {10.21105/joss.02004}, \href
  {https://ui.adsabs.harvard.edu/abs/2020JOSS....5.2004V} {5, 2004}

\bibitem[\protect\citeauthoryear{van~der Velden, Robert, Batten, Clauss,
  beskep, (Daniel)  \& Thyng}{van~der Velden et~al.}{2024}]{cmasher1.8.0}
van~der Velden E.,  Robert C.,  Batten A.,  Clauss C.,  beskep (Daniel) H.,
  Thyng K.,  2024, 1313e/CMasher: v1.8.0, \mn@doi{10.5281/zenodo.10677366},
  \url {https://doi.org/10.5281/zenodo.10677366}

\makeatother
\end{thebibliography}



\appendix

\section{Catalogue of accreted structures} \label{app:datatables}

In Table~\ref{tab:accretion-props} we present the complete catalogue of accretion events identified in Paper I, in addition to orbital properties determined in this work (Section~\ref{sec:fitting}).
The system IDs match the `accreted particle lists' in the Auriga public data release \citep{Grand:2024}.

\begin{table*}
\centering
\begin{tabular}{ccccccccccccc}
\hline
Halo & Level & ID & Morphology & $\log_{10}(M_\ast / \text{M}_\odot)$ & $f_\text{bound}$ & $r_\text{peri}$ & $r_\text{apo}$ & $t_\text{acc}$ & Distance & Preprocessed & Match ID \\
 &  &  &  &  &  & (kpc) & (kpc) & (Gyr) & (kpc)  &  \\
\hline
\multicolumn{10}{c}{$\cdots$} \\
6 & 3 & 176 & intact & 9.54 & 0.997 & 152.99 & -- & 1.94 & 199.79 & False & 232 \\
6 & 3 & 22562 & phase-mixed & 9.43 & 0.000 & 0.00 & 0.05 & 11.05 & -- & False & 433688 \\
6 & 3 & 144 & phase-mixed & 9.05 & 0.000 & 0.01 & 0.39 & 9.22 & -- & False & 102240 \\
6 & 3 & 2518 & phase-mixed & 9.04 & 0.000 & 0.01 & 0.79 & 8.90 & -- & False & 91904 \\
6 & 3 & 2333 & phase-mixed & 9.00 & 0.000 & 0.85 & 24.42 & 8.90 & -- & False & 3612 \\
6 & 3 & 151 & phase-mixed & 8.96 & 0.000 & 0.03 & 2.59 & 8.60 & -- & False & 184 \\
6 & 3 & 92 & phase-mixed & 8.78 & 0.000 & 0.01 & 0.39 & 9.52 & -- & False & 91897 \\
6 & 3 & 175 & intact & 8.40 & 0.990 & 73.79 & 225.01 & 6.07 & 129.62 & False & 231 \\
6 & 3 & 40792 & phase-mixed & 8.38 & 0.000 & 0.00 & 0.08 & 11.66 & -- & False & 660895 \\
6 & 3 & 2763 & stream & 7.97 & 0.410 & 15.87 & 165.67 & 6.40 & 37.60 & False & 3372 \\
6 & 3 & 2729 & stream & 7.73 & 0.001 & 27.64 & 55.02 & 8.90 & 44.65 & False & 102455 \\
6 & 3 & 13801 & phase-mixed & 7.42 & 0.000 & 0.02 & 1.20 & 9.22 & -- & False & -- \\
6 & 3 & 40886 & phase-mixed & 7.24 & 0.000 & 0.03 & 4.98 & 11.66 & -- & False & 484645 \\
6 & 3 & 33060 & phase-mixed & 7.16 & 0.000 & 11.64 & 97.19 & 8.60 & -- & True & 403733 \\
6 & 3 & 365 & intact & 7.03 & 0.998 & 57.14 & -- & 2.25 & 217.36 & False & 3105 \\
6 & 3 & 13857 & stream & 6.98 & 0.134 & 22.41 & 99.02 & 10.75 & 35.11 & False & 392843 \\
6 & 3 & 82 & phase-mixed & 6.96 & 0.000 & 0.95 & 14.15 & 11.93 & -- & False & 151 \\
6 & 3 & 450129 & phase-mixed & 6.93 & 0.000 & 0.04 & 1.54 & 12.18 & -- & False & 1346443 \\
6 & 3 & 441235 & phase-mixed & 6.81 & 0.000 & 0.01 & 1.04 & 11.66 & -- & False & 1375567 \\
6 & 3 & 1022 & stream & 6.80 & 0.544 & 18.15 & 334.69 & 8.28 & 197.17 & True & 7306 \\
6 & 3 & 40713 & phase-mixed & 6.78 & 0.000 & 15.91 & 59.88 & 11.36 & -- & False & 608072 \\
6 & 3 & 86344 & stream & 6.70 & 0.000 & 26.24 & 156.81 & 8.60 & -- & True & 423599 \\
6 & 3 & 54423 & intact & 6.52 & 1.000 & -- & -- & -- & 270.26 & False & 8695 \\
6 & 3 & 4317 & stream & 6.49 & 0.921 & 26.87 & 364.90 & 7.99 & 97.46 & False & 96349 \\
6 & 3 & 439497 & stream & 6.47 & 0.097 & -- & 57.75 & 11.36 & 32.36 & False & 608129 \\
6 & 3 & 66116 & phase-mixed & 6.46 & 0.000 & 0.34 & 15.55 & 11.66 & -- & False & 1969087 \\
6 & 3 & 86785 & phase-mixed & 6.45 & 0.000 & 0.52 & 17.84 & 8.60 & -- & True & 510516 \\
6 & 3 & 450183 & phase-mixed & 6.42 & 0.000 & 0.17 & 5.46 & 12.56 & -- & False & 1346695 \\
6 & 3 & 60914 & phase-mixed & 6.41 & 0.000 & 0.23 & 4.67 & 12.18 & -- & False & 834323 \\
6 & 3 & 450996 & phase-mixed & 6.40 & 0.000 & 0.12 & 2.70 & 12.56 & -- & False & 1346807 \\
6 & 3 & 820096 & phase-mixed & 6.35 & 0.000 & 0.06 & 3.10 & 11.93 & -- & False & 661067 \\
6 & 3 & 119824 & stream & 6.21 & 0.000 & 10.94 & 169.01 & 8.90 & -- & True & 510560 \\
6 & 3 & 401238 & phase-mixed & 6.19 & 0.000 & 0.95 & 38.04 & 8.90 & -- & True & 485312 \\
6 & 3 & 818755 & phase-mixed & 6.15 & 0.000 & 18.64 & 47.26 & 11.36 & -- & False & 786494 \\
6 & 3 & 22636 & phase-mixed & 6.14 & 0.000 & 1.51 & 20.92 & 11.05 & -- & False & 709923 \\
6 & 3 & 83559 & phase-mixed & 6.09 & 0.000 & 29.49 & 117.54 & 9.52 & -- & True & 485601 \\
6 & 3 & 820255 & phase-mixed & 6.07 & 0.000 & 1.47 & 9.01 & 11.93 & -- & False & 7675517 \\
6 & 3 & 86573 & phase-mixed & 6.00 & 0.000 & 1.48 & 29.02 & 8.60 & -- & True & 510832 \\
6 & 3 & 98912 & stream & 5.98 & 0.525 & 25.77 & 235.18 & 10.13 & 206.71 & False & 423105 \\
6 & 3 & 22641 & stream & 5.96 & 0.000 & 100.50 & 229.46 & 11.05 & -- & False & 480459 \\
6 & 3 & 105073 & intact & 5.96 & 0.978 & 60.45 & 297.28 & 6.07 & 177.53 & False & 527547 \\
6 & 3 & 451182 & phase-mixed & 5.94 & 0.000 & 0.04 & 1.17 & 12.56 & -- & False & 3975982 \\
6 & 3 & 25426 & stream & 5.84 & 0.067 & 52.84 & 100.88 & 9.22 & 65.77 & False & 439605 \\
6 & 3 & 450937 & phase-mixed & 5.83 & 0.000 & 0.59 & 9.11 & 12.56 & -- & False & 1346868 \\
6 & 3 & 466589 & phase-mixed & 5.80 & 0.000 & 0.34 & 6.34 & 12.56 & -- & False & 3971198 \\
6 & 3 & 461836 & phase-mixed & 5.74 & 0.000 & 0.00 & 0.51 & 13.07 & -- & False & 4028766 \\
6 & 3 & 441640 & phase-mixed & 5.71 & 0.000 & 0.03 & 4.03 & 11.66 & -- & False & 661038 \\
6 & 3 & 461788 & phase-mixed & 5.71 & 0.000 & 0.03 & 1.09 & 12.85 & -- & False & 140 \\
6 & 3 & 60987 & phase-mixed & 5.70 & 0.000 & 0.06 & 6.22 & 12.85 & -- & False & 1735450 \\
6 & 3 & 52829 & phase-mixed & 5.69 & 0.000 & 0.83 & 14.32 & 9.52 & -- & True & 1719764 \\
\multicolumn{10}{c}{$\cdots$} \\
\hline
\end{tabular}

    \caption{
    Catalogue of accretion events and their properties analysed in this work, sorted by level, then halo number, then stellar mass.
    For brevity, we only show systems for the level 3 run of Au-6 in this manuscript.
    We provide the halo number, resolution level, and system ID that in combination uniquely identify an object (IDs alone are not guaranteed to be unique); morphological classification; total stellar mass ($M_\ast$), including bound progenitor if still present at $z=0$; fraction of stellar mass bound to the progenitor ($f_\text{bound}$); pericentre ($r_\text{peri}$); apocentre ($r_\text{apo}$); accretion time ($t_\text{acc}$) defined as first crossing of the host's $R_\text{200c}$; distance from the host at the present day for accretions with a bound progenitor; and the matched ID of the same object at one resolution level higher.
    A machine-readable table with the full catalogue is available as supplementary material.
    }
    \label{tab:accretion-props}
\end{table*}


\bsp	
\label{lastpage}
\end{document}